%% file: main.tex
\documentclass[prl,reprint,aps,amsmath,amssymb,superscriptaddress,floatfix,secnumarabic]{revtex4-2}

\input{packages}
\makeatletter
\@booleantrue\acknowledgments@sw
\def\@bibdataout@aps{%
  \immediate\write\@bibdataout{%
    @CONTROL{%
      apsrev42Control%
      \longbibliography@sw{%
        ,author="48",editor="1",pages="0",title="0",year="1"%
      }{%
        ,author="48",editor="1",pages="0",title="",year="1"%
      }%
    }%
  }%
  \if@filesw
    \immediate\write\@auxout{\string\citation{apsrev42Control}}%
  \fi
}
\hypersetup{
  colorlinks=true,
  citecolor=blue,
  urlcolor=blue,
  linkcolor=red,
  breaklinks=true
}
\begin{document}

\setcounter{secnumdepth}{1}

\title{Entanglement (1+2) QED in a double layer of Dirac Materials}

\input{authors}	
	
\begin{abstract}
We investigate the momentum-space entanglement between two Dirac quasiparticles in a double-layer honeycomb lattice coupled via a planar electromagnetic cavity. We model the low-energy excitations as massive Dirac fermions in $(1+2)$ dimensions and derive the Bethe–Salpeter equation using the ladder approximation. We use a Born-level approximation around a free two-body quasiparticle state, where the interaction is mediated by the cavity photon propagator. From the reduced sublattice density matrix, we compute a momentum-resolved von Neumann entropy.
Within the perturbatively controlled regime, the entropy remains small, while phenomenological self-energy dressing drives a crossover to strong enhancement of the entanglement entropy. Stationary entanglement is obtained only when the quasiparticle coherence time exceeds the photon propagation time between the layers. The maximum-entropy regime appears to be a viable method for achieving Bell-like states. These results demonstrate how self-energy renormalization, virtual particle exchange, and spinor geometry combine to reshape the entanglement landscape of Dirac materials.
\end{abstract}
	
\maketitle

\section{Introduction}\label{sec:introduction}
Quantum information theory is a rapidly developing field at the intersection of quantum mechanics and computer science. At its core, the concept of entanglement is conceived as a physical resource that can be harvested~\cite{reznik2}, enabling quantum teleportation~\cite{bennet93}, quantum cryptography~\cite{Ekert91}, and quantum computation~\cite{jozsa2003role}. The generation of entanglement between two two-level detector systems has been studied extensively through their interaction with the ground state of a quantum field~\cite{resnik,valentini}: because this ground state carries spatial correlations, these correlations can be transferred to the detectors, thereby generating entanglement between otherwise non-interacting subsystems. More recently, the same question has been addressed for pairs of massive charged particles whose correlations are mediated by the exchange of virtual particles~\cite{horodecki,perche}. The converse situation, in which the quantum fields themselves play the role of the entangled parties, has received comparatively less attention~\cite{pozas}, partly because a quantum field permeates all of space, and the natural bipartition is between excitations rather than between localized regions.

Beyond this foundational setting, graphene-family materials are increasingly being discussed as active components of quantum-information hardware. Proposed implementations span room-temperature logic gates based on ballistic Dirac transport~\cite{dragoman2016logic}, continuous-variable teleportation in plasmonic graphene waveguides, and local-area teleportation networks~\cite{asjad2021teleport,asjad2022network}, graphene-based qubits in communication architectures~\cite{wu2012graphenequbits}, and decoherence-robust correlation indicators such as uncertainty-induced non-locality~\cite{mohamed2022nonlocal}. Together, these proposals motivate viewing Dirac materials not only as analog simulators of relativistic physics but also as prospective building blocks for scalable quantum technologies~\cite{Cayssol2013, Calafell2019}. The present work contributes a complementary ingredient: rather than postulating a specific device operation, we identify a cavity-mediated regime in which a Bethe-Salpeter/QED description yields strong interlayer entanglement, providing a concrete roadmap toward Bell-type resources once the corresponding homogeneous equation is solved to isolate genuinely bound entangled states.

It is well-established that electrons in graphene behave as massless relativistic Dirac particles~\cite{peres-guinea}, which allows for the application of the high-energy physics mathematical framework, such as quantum field theory (QFT)~\cite{peskin} and quantum electrodynamics (QED), to the understanding of condensed matter phenomena~\cite{marino,lamata,pacho}. However, there is comparatively little literature on related materials with honeycomb structures that exhibit spin-orbit interactions as a result of lattice corrugations~\cite{kanemele}, where the coupling manifests as a second-neighbor interaction and contributes a mass term to the diagonal of the tight-binding Hamiltonian. The motivation for a QED perspective is particularly natural in layered two-dimensional systems: the electronic fields in distinct layers do not share coordinates, so any interlayer correlations must arise from the exchange of virtual photons. A planar microcavity provides a clean, tunable electromagnetic environment in which this exchange can be controlled through cavity geometry and mode structure~\cite{li,keller}, allowing entanglement to be quantified directly from the two-particle amplitude~\cite{lamata} while avoiding point-like interactions and highlighting non-local correlations~\cite{byrnes,Ysun,facu1,facu2}.

In this context, two-dimensional electrons in honeycomb lattices such as silicene~\cite{spen}, germanene~\cite{Zhang2020,bian}, and related materials~\cite{roldan} can be regarded as massive Dirac particles, with the mass given by the spin-orbit coupling rather than the electron rest mass. We employ a (1+2)-dimensional QFT formalism~\cite{jsa2018,jsa1,jsaA}---an analogous framework applicable within the correct energy regimes, with the Fermi velocity replacing the speed of light~\cite{liu,yao,ezawa}---to derive the Bethe-Salpeter equation for a two-body Dirac-quasiparticle state~\cite{greiner} in a double-layer honeycomb lattice coupled to a planar cavity. We obtain the single-photon exchange kernel in momentum space, evaluate the corresponding first-order correction to the two-body wave function, and compute a momentum-resolved von Neumann entropy~\cite{vonneumann} from the reduced sublattice (pseudospin) density matrix, analyzing how a phenomenological quasiparticle self-energy parameter modifies this quantity.

This framework connects with several recent developments. In previous works, we showed that virtual-photon-mediated correlations can be tuned in twisted double-layer graphene and in buckled honeycomb lattices embedded in microcavities, where geometric parameters such as the twist angle and the layer positions inside the cavity act as control knobs for the harvested correlations~\cite{facu1,facu2}. Zhang \textit{et al.}\ introduced a non-Hermitian Bethe--Salpeter equation for open quantum systems, providing a systematic route to include dissipation, causality, and nonequilibrium spectral effects beyond a purely Hermitian description~\cite{zhang2025nhbse}. In a conceptually related scattering setting, Kulig \textit{et al.}\ showed that altermagnetic spin-dependent scattering can be engineered to generate Bell states with high fidelity~\cite{kulig2025altermagnet}. These developments place the present ladder-Bethe--Salpeter treatment within a broader program aimed at engineering entanglement in solid-state platforms through geometry, dissipation, and spinor-resolved scattering.

Our results highlight that, away from the resonantly enhanced region, cavity-mediated entanglement remains in the perturbative low-entanglement regime expected for a weakly interacting process. Phenomenological self-energy dressing drives a crossover toward strong enhancement, which persists over a broad momentum window and is strongly suppressed for near-collinear kinematics or for coherence times shorter than the photon travel time between the layers. The paper is organized as follows. Section~\ref{sec:theoretical model} introduces the effective Dirac model and the cavity photon propagator. Section~\ref{sec:theoretical_framework} develops the two-particle QED framework. Section~\ref{sec:results} presents the entanglement results, followed by a discussion in Section~\ref{sec:discussion} and conclusions in Section~\ref{sec:conclusions}. Technical derivations and explicit wave-function components are collected in Appendices~\ref{app:bethe_salpeter} and~\ref{app:numerical_construction}.

\begin{figure}[t]
    \centering
    \includegraphics[width=0.92\linewidth]{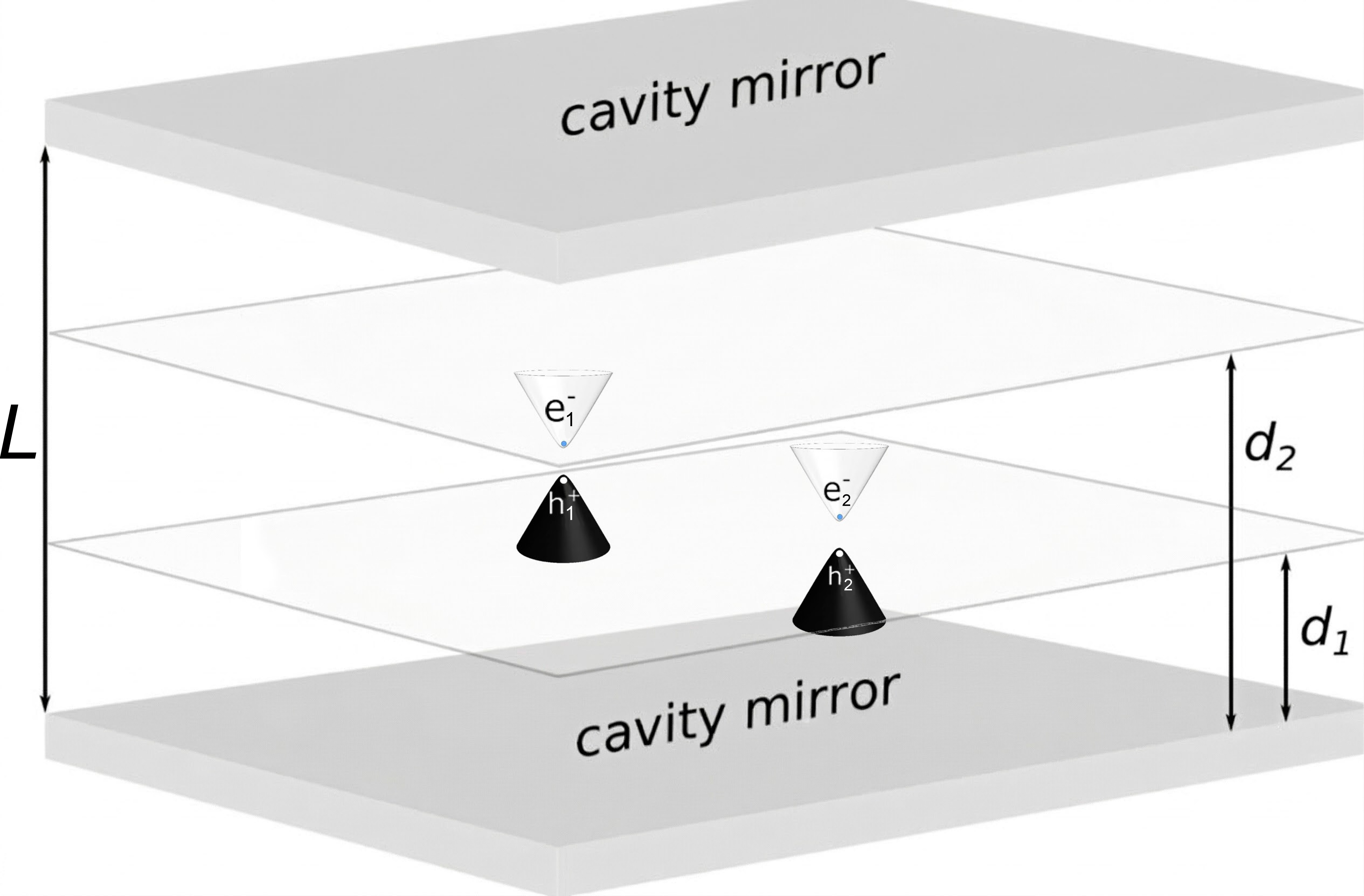}\par\vspace{0.6em}
    \includegraphics[width=0.92\linewidth]{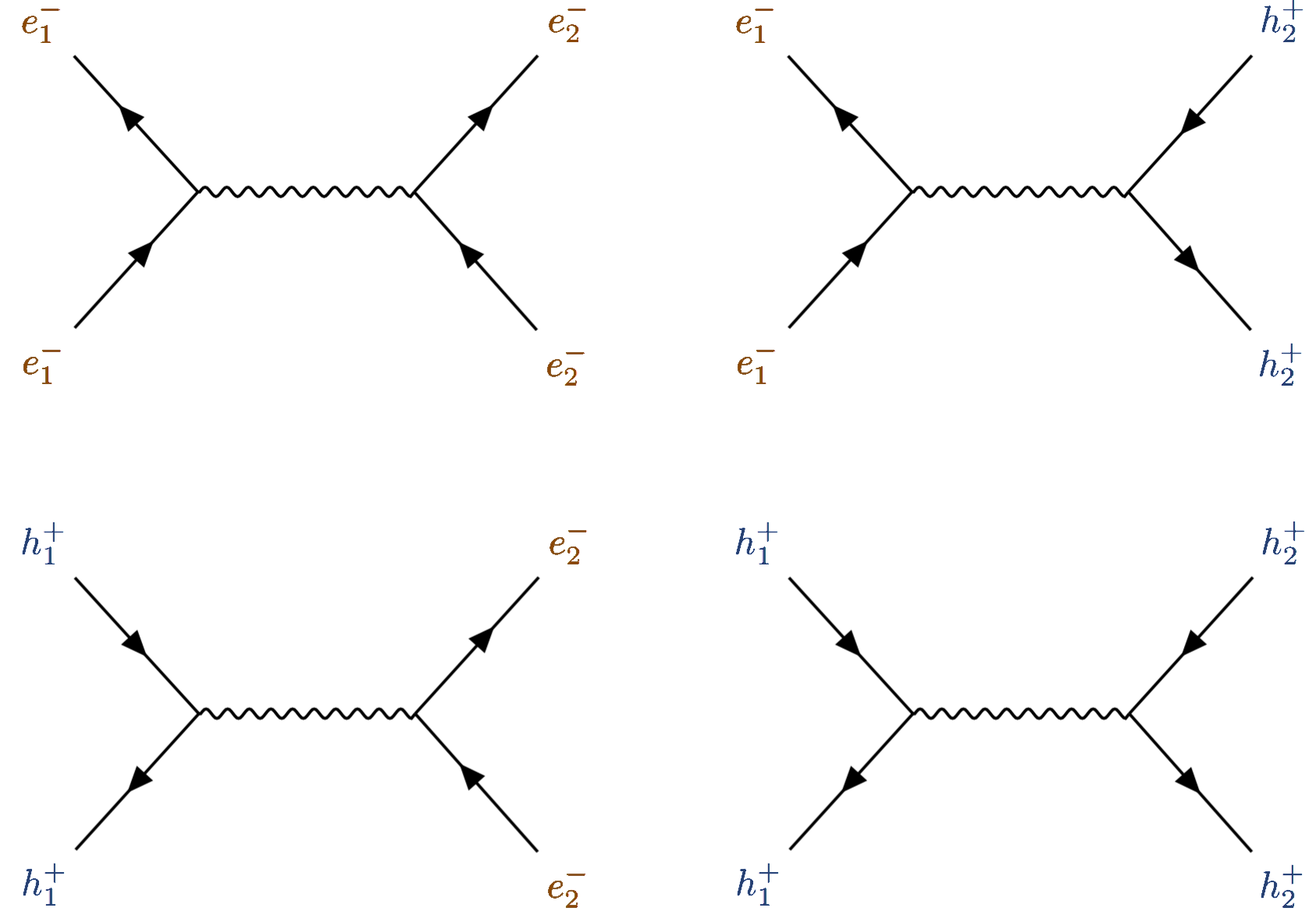}
    \caption{Schematic representation of the theoretical setup used in this work. \textit{Top}: gapped Dirac-cone structure associated with the honeycomb layers. \textit{Bottom}: representative cavity-mediated interlayer interaction channels used in the calculation. The diagrams summarize the four sectors entering the numerical construction of the two-body state---electron--electron (ee), electron--hole (eh), hole--electron (he), and hole--hole (hh)---all connected by virtual-photon exchange between the layers.}
    \label{fig:setup_feynman}
\end{figure} 

\section{Theoretical Model}\label{sec:theoretical model}
 Our starting point is the tight-binding Hamiltonian of the honeycomb lattice, which leads us to an effective theory of Dirac fermions \cite{kristi,spen,peres-guinea}
 \begin{equation}\label{Ham}
	H_{s,\eta}=  v_F \boldsymbol{\sigma} \cdot \vb{p}_\eta +\sigma_{z}\Delta_{s,\eta},
\end{equation}
where $s,\eta=\pm$ are the spin and valley indices. \( \boldsymbol{\sigma} = \left( \sigma_{x}, \sigma_{y} \right) \) and \( \sigma_{z} \) are the Pauli matrices that act within each sublattice space. Since the two valleys are separated by a large momentum gap, interval scattering is generally negligible at low energies \cite{par,zare1}. Therefore, in our analysis, we will only focus on one valley and, for convenience, we will take $\eta=1$. The Fermi velocity \( v_{F} \) is specific to the material; for example, \( v_{F} = 5.5 \times 10^{5} \) $m/s$ in silicene \cite{liu}. \( \Delta_{s} = s\lambda_{so} - \ell E_{z} \), where \( \lambda_{so} \) is the spin-orbit coupling constant, \( \ell \) is the distance between the sublattices \( A \) and \( B \), and \( E_{z} \) is a perpendicular electric field that can tune the band gap. In the following discussion, we simplify the setup by assuming the absence of an external perpendicular electric field. The Hamiltonian of the Dirac material can then be written as
 \begin{equation}
    H_s = \left(\begin{array}{cc}
     s \lambda_{so}   &  v_F (p_x - i p_y)   \\
         v_F (p_x + i p_y) & -s \lambda_{so}
    \end{array} \right).
 \end{equation}
The eigenvectors of this Hamiltonian can be written as
\begin{equation}\label{wave}
    \psi_{\nu,s}(t,\vb{r}) = \int \frac{d^2 p}{(2\pi)^2} \, u_{\nu,s}(\vb{p}) \, e^{-i \nu\left(\frac{E(\vb{p})}{\hbar}  t-\vb{p}\cdot\vb{r}\right)},
\end{equation}
where  $\nu=\pm$ is the band index (the analog to the positive and negative energy in relativistic particles), and $u_{\nu,s}$ represents spinors that depend on the band index.  Near the Dirac point, the valence-band branch ($\nu=-$) is interpreted as a hole excitation, so we adopt the standard QFT notation $u_{+,s}(\vb{p})\equiv u_s(\vb{p})$ and $u_{-,s}(\vb{p})\equiv v_s(\vb{p})$. Thus, for each layer, we write
\begin{equation}\label{ket e}
	u_{+s} (\vb{p})=u_{s} (\vb{p})= \frac{1}{\sqrt{2 S \left[ 1+\left( \chi^{+}_{s}\right)^2  \right] }} \left(\begin{array}{c}
		1 \\ \chi^{+}_{s} e^{i \phi_{\vb{p}}}
	\end{array}\right),
\end{equation}
\begin{equation}\label{ket hole}
	u_{-s} (\vb{p})=v_{s} (\vb{p})= \frac{1}{\sqrt{2 S \left[ 1+\left( \chi^{-}_{s}\right)^2  \right] }} \left(\begin{array}{c}
		1 \\ \chi^{-}_{s} e^{i \phi_{\vb{p}}}
	\end{array}\right),
\end{equation}
where $ \phi_{\vb{p}}= \operatorname{atan2}\!\left( p_y,p_x \right) \in (-\pi,\pi] $ and the remaining terms read
\[\chi^{\pm}_{s}(\vb{p})= \dfrac{ v_F \abs{\vb{p}}}{\pm E (\vb{p}) + \Delta_{s}}= \dfrac{\pm E(\vb{p})- \Delta_{s}}{v_F \abs{\vb{p}}}\]
\[E(\vb{p}) =  \sqrt{ (v_F \abs{\vb{p}})^2 + \lambda_{so}^2}.   \]

The linear dispersion relation around the Dirac points results in a double cone with a gap determined by the mass and is equivalent to the relativistic energy of a particle. Therefore, the electrons close to the Dirac points effectively behave as Dirac fermions, but the role of the speed of light is taken by the Fermi speed $ v_F $.

It is now convenient to adopt a relativistic notation for spacetime by making the following definitions
\[
\begin{aligned}
	x^\mu & =\left(v_F t, x, y\right), \\
	g_{\mu \nu} & =\operatorname{diag}(1,-1,-1), \\
	\partial_\mu & =\frac{\partial}{\partial x^\mu}=\left(\frac{1}{v_F} \frac{\partial}{\partial t}, \nabla\right), \\
	\gamma^\mu & =\sigma_z\left(1, \sigma_x, \sigma_y\right)=\left(\gamma^0, \boldsymbol{\gamma} \right),
\end{aligned}
\]
where the index $\mu$ goes from 0 to 2 , with component 0 always associated with time and 1, 2 with spatial coordinates, $x^\mu$ is the $\mu-$th component of the vector of coordinates in space-time in $1+2$ dimensions, $g_{\mu \nu}=g^{\mu \nu}$ is the metric tensor in Minkowski space, `diag'  denotes a diagonal matrix with diagonal elements given in parentheses, $\partial_\mu$ is the differential operator and $\gamma^\mu=\left(\gamma^0, \gamma^1, \gamma^2\right)$ are the Dirac matrices, which must satisfy the algebra
\begin{equation*}
\left\{\gamma^\mu, \gamma^\nu\right\}=\gamma^\mu \gamma^\nu+\gamma^\nu \gamma^\mu=2 g^{\mu \nu} \mathbf{1},
\end{equation*}
where $\mathbf{1}$ is the identity matrix with the dimensionality of the Dirac matrices.

Using Einstein's convention, it can be shown that the Dirac equation can be rewritten in the compact form:
\begin{equation}\label{Dirac}
	\left( i \hbar \,\partial \!\!\!/ -  s\,m \right)\psi_s=0,
\end{equation}
where we introduced the effective Dirac mass parameter
\(m\equiv \lambda_{so}/v_F\), so that the on-shell condition reads
\(p^\mu p_\mu=m^2\) when the temporal component is chosen as \(p^0=E/v_F=\pm\sqrt{p^2+m^2}\), with $p=\abs{\vb{p}}$.

The canonical field quantization implies replacing the ``classical'' wave functions $\psi_{s}$ by the field operators:
	\begin{align}\label{psis}
		&\hat{\psi}_{s}(x)  = \int \frac{\mathrm{d}^2 p}{(2 \pi)^2} \left(\hat{a}_{\vb{p}}^{s}  u_{s}(\vb{p})  \mathrm{e}^{-\mathrm{i} p \cdot x}+\hat{b}_{\vb{p}}^{s\dagger}  v_{ s}(\vb{p}) \mathrm{e}^{\mathrm{i} p \cdot x}\right), \\
		&\hat{\bar{\psi}}_{s}(x)  =\int \frac{\mathrm{d}^2 p}{(2 \pi)^2} \left(\hat{a}_{\vb{p}}^{s\dagger} \tilde{u}_{ s}(\vb{p})  \mathrm{e}^{\mathrm{i} p \cdot x}+\hat{b}_{\vb{p}}^{s} \tilde{v}_{ s}(\vb{p})  \mathrm{e}^{-\mathrm{i} p \cdot x}\right).
	\end{align}

Then, one defines the time-ordered contraction (or simply the contraction) of two operators as:
\[
\left\langle 0\left |T\hat{\psi}_{s} (x) \hat{\overline{\psi}}_{s'}(x')\right| 0\right\rangle= i S_{\mathrm{F}ss'}\left(x-x'\right).
\]
Therefore for spin- $\frac{1}{2}$ particles, the propagator reads
\begin{align}\label{eq:S}
S_{\mathrm{F}s}(x-x')&=
\int \frac{\mathrm{d}^3 p}{(2 \pi)^3} \: \frac{(\not p  + s \, m)}{ p^2-m^2+\mathrm{i} \epsilon} \:\mathrm{e}^{-\mathrm{i} p \cdot(x-x')}  . 
\end{align}

For simplicity, we restrict the numerical analysis to one spin sector (\(s=+1\)). This is a modeling choice (not a strict equivalence), and the opposite spin sector \((s=-1)\) can be treated analogously in an extended calculation.

\subsection{Photon Propagator}\label{subsec:Photons}
Having established the two-particle framework, we now introduce the interaction responsible for mediating correlations between the electrons. In the present system, this interaction is provided by the quantized electromagnetic field confined within a planar microcavity \cite{kib}. For the internal photon lines, we adopt a covariant formulation in the Feynman gauge, which yields a compact tensor structure for the propagator.

The electromagnetic vector potential takes the form
\begin{align}\label{Akib}
	\mathbf{A}_{i}(x,y,z=d_{i},t)&
	=\int \frac{\mathrm{d}^2 q}{(2 \pi)^2} \sum_n \frac{\zeta}{\sqrt{2\omega_{n,\mathbf{q}}}}\sin \left( \frac{n \pi d_{i}}{L}\right) \nonumber\\
	&\times \sum_{\lambda=\pm} \Big(\bm{\epsilon}_{\lambda} a_{n\mathbf{q}\lambda}e^{-i(\omega_{n \mathbf{q} } t -q_x x -q_y y)} \nonumber \\& \qquad
	+\bm{\epsilon}_{\lambda}^{\ast}a_{n\mathbf{q}\lambda}^{\dagger }e^{i(\omega_{n\mathbf{q} } t -q_x x -q_y y)}\Big) ,
\end{align}
where $\zeta=\sqrt{\hbar/\varepsilon_0 \varepsilon_r L}$ and $\omega_{n,\vb{q}}=c\sqrt{|\vb{q}|^2+\left(\frac{n\pi}{L}\right)^2}$ are the cavity mode amplitudes and dispersions, respectively. We assume periodic boundary conditions in the in-plane directions with quantization area $S$; the corresponding normalization factors are absorbed into the field amplitude. In the numerical implementation, we use natural units ($\hbar=c=1$) after converting all dimensional quantities to a consistent eV--nm system; consequently, $\zeta$ and the effective coupling are handled as rescaled parameters in that convention.

Within this quantized description, the interaction between the electrons is mediated by the exchange of virtual cavity photons. The corresponding photon propagator is obtained from the time-ordered contraction of two vector potential operators,
\[
\wick[positions={-1}]{
\c{\hat{A}}^\mu(x)\c{\hat{A}}^\nu(x')
=
\left\langle 0 \left|
T\!\left(
\hat{A}^\mu(x)\hat{A}^\nu(x')
\right)
\right| 0 \right\rangle
=
\mathrm{i} D_{\mathrm{F}}^{\mu\nu}(x-x').
}
\]

Combining these modeling choices, the interlayer photon propagator between electrons located at layers \(1\) and \(2\) is written in a compact form.
\begin{align}
D^{\mu \nu}_{\mathrm{F}}& (x_1 - x'_2)
=
-\zeta^2
\sum_n
\sin\!\left(\frac{n\pi d_1}{L}\right)
\sin\!\left(\frac{n\pi d_2}{L}\right)
\nonumber\\
&\quad \times
\int \frac{\mathrm{d}^3 q}{(2\pi)^3}
\mathrm{e}^{-\mathrm{i} q\cdot(x_1-x'_2)}
\frac{g^{\mu\nu}}
{q^2-\left(\frac{n\pi}{L}\right)^2+\mathrm{i}\epsilon},
\label{photonpro}
\end{align}
where we integrate over the off-shell three-momentum \(q^\mu=(q^0,\vb{q})\) with \(q^2\equiv (q^0)^2-|\vb q|^2\). The cavity confinement enters through the discrete transverse momentum \(q_z=n\pi/L\), which shifts the photon denominator by \((n\pi/L)^2\).

\section{Two-Body QED Framework} \label{sec:theoretical_framework}

To investigate the momentum-space entanglement properties of the two-body quasiparticle system, we consider the interacting two-body wave function. We define the Bethe-Salpeter amplitude $\varphi_{12}(x_1, x_2)$ as the vacuum expectation value of the time-ordered product of the field operators. In spinor space, we treat $\varphi_{12}$ as a bispinor object (equivalently, a matrix with one spinor index per particle), so inverse propagators act on the corresponding left/right indices. In momentum space, we work with the unamputated amplitude, which satisfies the Bethe-Salpeter equation:
\begin{align}\label{BS}
(\slashed{p}_1 - m_1)\varphi_{12} (p_1,p_2) (\slashed{p}_2 - m_2) \nonumber\\
= \int d^3 p'_1 d^3 p'_2 \bar{K}^{ab} (p_1, p_2 , p'_1 , p'_2) \varphi(p'_1,p'_2),
\end{align}
where $\bar{K}^{ab}$ is the interaction kernel encoding the cavity-mediated electromagnetic coupling. Here, the RHS acts as a source term evaluated on the free two-body state $\varphi^{(0)}$.

In Eq.~\eqref{BS}, the terms $(\slashed{p}_{1,2} - m_{1,2})$ represent the inverse of the bare electron propagators $S_0^{-1}(p)$. To incorporate self-energy effects without solving the full Dyson–Schwinger equations, we upgrade the bare propagators to dressed quasiparticle propagators $S_{\mathrm{eff}}(p)$, adopting the on-shell quasiparticle approximation in which the self-energy $\Sigma(p)$ is treated as a local phenomenological constant. This is justified by two observations. First, the momentum range explored in our results is narrow and lies well below the spin--orbit gap scale, $m=\lambda_{\mathrm{so}}/v_F \approx 2.13$~eV, so the dressed quasiparticle dispersion is approximately linear and the mass renormalization is nearly uniform across the momentum grid. Second, in the related context of massless reduced quantum electrodynamics (RQED$_{4,3}$) applied to graphene, Kotikov and Teber~\cite{kotikov2014} showed at two-loop order that the anomalous scaling dimension of the fermion field is independent of the external momentum (see Eq.~(84) of \cite{kotikov2014}). While our system is gapped and confined to a planar cavity, and is therefore not directly described by RQED$_{4,3}$, this result offers a heuristic analogy suggesting that self-energy corrections in low-dimensional Dirac systems may vary slowly with momentum. We adopt the constant-$\Sigma$ approximation as a phenomenological working hypothesis. Concretely, $\mathrm{Re}(\Sigma)=\Sigma'$ is the static mass renormalization of the quasiparticle, while $\mathrm{Im}(\Sigma) = \Gamma > 0$ introduces a finite coherence time (decay rate) 
\begin{equation}
\tau_{\mathrm{coh}}=\frac{1}{\mathrm{Im}\left(\Sigma\right)},
\end{equation}
regularizing the bound-state pole. Mathematically, this corresponds to the substitution in the LHS of Eq.~\eqref{BS}:
\begin{equation}
    (\slashed{p} - m) \longrightarrow S_{eff}^{-1}(p) \approx (\slashed p - m - \Sigma).
\end{equation}
The interaction kernel is then treated in the ladder approximation (single-photon exchange):
\begin{align} \label{kernel}
K^{12} (r'_1 , r'_2; r_1, r_2 ) \approx (-ie)^2 \gamma_\mu^a D_F^{\mu \nu}(r_1,r_2)\gamma_\nu^b \nonumber\\
\times \delta^{(3)}(r'_1-r_1 ) \delta^{(3)}(r'_2 - r_2 ),
\end{align}
where $D_F^{\mu\nu}$ is the photon propagator of Section~\ref{sec:theoretical model}.

Finally, projecting the dressed propagators onto the mass shell (as detailed in Appendix~\ref{app:bs-frequency}), Eq.~\eqref{BS} reduces to the effective eigenvalue equation solved in this work:
\begin{align}\label{final}
(\slashed{p}_1 - m_1 - \Sigma_1)\varphi_{12} (p_1,p_2) (\slashed{p}_2 - m_2 - \Sigma_2) \nonumber \\
= -e^2 \int \dfrac{d^3 q}{(2\pi)^3} \; \mathcal{D}(q) \gamma_\mu^a \gamma_b^\mu\;\varphi_{12}(p_1 - q ,p_2 +q),
\end{align}
where $\mathcal{D}(q)$ is the effective cavity photon propagator.

In the present manuscript, we do not solve Eq.~\eqref{final} self-consistently. Instead, we evaluate its first iterative correction, i.e., a Born-level treatment around the free two-body quasiparticle state. The derivation of the kernel, the intermediate integrations, and the resulting spinor structure are collected in Appendix~\ref{app:bethe_salpeter}, while the numerical construction of the total state is summarized in Appendix~\ref{app:numerical_construction}. In this representation, the two-body wave function can be expressed in the product basis
\(\{|A{1},A{2}\rangle, |A{1},B{2}\rangle, |B{1},A{2}\rangle, |B{1},B{2}\rangle\}\).
Here $A/B$ denotes the two sublattices (pseudospin), and the numbers 1 and 2 correspond to the layer at positions $d_1$ and $d_2$. 

\subsection{Reduced Density Matrix and Entanglement Entropy}
\label{subsec:rhoA}
To quantify the bipartite entanglement between the internal degrees of freedom (sublattice pseudospins) associated with the electrons, we adopt a momentum-resolved approach. Specifically, we compute the entanglement of the \textit{conditional state} $|\psi_{\mathbf{p}}\rangle$ for a fixed kinematic configuration $\mathbf{p}=\{p_1, p_2\}$. This corresponds to analyzing the correlations between the sublattice indices, given that the electrons are detected with specific momenta.

The full two-body wave function obtained from the Bethe-Salpeter equation can be projected onto the spinor subspace spanned by the product basis $\{|A1,A2\rangle,|A1,B2\rangle,|B1,A2\rangle,|B1,B2\rangle\}$. For a fixed momentum configuration, the conditional state is written as:
\begin{align} |\psi_{\mathbf{p}}\rangle = \mathcal{N}_{\mathbf{p}} \left( c_{AA}|A{1},A{2}\rangle + c_{AB}|A{1},B{2}\rangle \right.\nonumber\\+ c_{BA}|B{1},A{2}\rangle + c_{BB}|B{1},B{2}\rangle \left.\right), \label{eq:psi_expansion} \end{align} where the coefficients $c_{ij}$ correspond to the four wave function components ($\varphi_{AA}, \varphi_{AB}, \dots$) derived in Appendix~\ref{app:bs-spinor}. Appendices~\ref{app:num-first-order} and~\ref{app:num-superposition} explain how these channel contributions are obtained and assembled numerically into the total state used below. Here, $\mathcal{N}_{\mathbf{p}}$ is a local normalization constant ensuring $\langle \psi_{\mathbf{p}} | \psi_{\mathbf{p}} \rangle = 1$. We emphasize that this local normalization renders $S_1(\mathbf{p})$ a conditional quantity: it measures the pseudospin entanglement of the two-body state given that the quasiparticles are detected with momenta $(p_1,p_2)$, and does not carry information about the probabilistic weight of that kinematic configuration. A momentum-weighted global measure would require integrating $S_1$ against the full two-body momentum distribution and is left for future work.

The reduced density matrix $\rho_{1}(\mathbf{p})$ is obtained by tracing out the degrees of freedom of subsystem 2 (layer 2) from this conditional state:
\begin{equation}
    \rho_{1}(\mathbf{p}) = \text{Tr}_{2}(|\psi_{\mathbf{p}}\rangle\langle\psi_{\mathbf{p}}|) = \sum_{j=A,B}\langle j2|\psi_{\mathbf{p}}\rangle\langle\psi_{\mathbf{p}}|j2\rangle.
\end{equation}
By construction, $\text{Tr}(\rho_1)=1$. We emphasize that this procedure defines an  entanglement entropy measure at fixed momenta: any global multiplicative factor in the scattering amplitude cancels in the local normalization. Consequently, this quantity is valid for fixed kinematics.

The matrix elements of $\rho_1$ are given by:
\begin{align}
    \rho_{1}(0,0) &= |c_{AA}|^2 + |c_{AB}|^2, \nonumber \\
    \rho_{1}(1,1) &= |c_{BA}|^2 + |c_{BB}|^2, \nonumber \\
    \rho_{1}(0,1) &= c_{AA}c_{BA}^* + c_{AB}c_{BB}^*, \nonumber \\
    \rho_{1}(1,0) &= c_{BA}c_{AA}^* + c_{BB}c_{AB}^*.
\end{align}
These matrix elements are then used to compute the entanglement entropy.

For each configuration, the eigenvalues $\{\nu_i\}_{i=1,2}$ of $\rho_1(\mathbf{p})$ are obtained, and the bipartite entanglement entropy is computed from the von Neumann expression
\begin{equation}
S_1(\mathbf{p}) = -\sum_{i=1}^2 \nu_i \log \nu_i .
\label{eq:entropy}
\end{equation}
In the numerical implementation, the two-body state is normalized as $\Psi_\mathrm{tot} \to \Psi_\mathrm{tot}/\|\Psi_\mathrm{tot}\|$ prior to constructing $\rho_1$, which enforces $\mathrm{Tr}(\rho_1)=1$ analytically. The matrix is then symmetrized as $\rho_1 \to \tfrac{1}{2}(\rho_1+\rho_1^\dagger)$ to remove floating-point asymmetry. The eigenvalues are computed and renormalized by their sum to restore the unit-trace condition. This procedure yields the entanglement profile $S_1(p_1,p_2,\phi_1,\phi_2)$ over the full momentum grid, forming the basis of the analysis presented in Sec.~\ref{sec:results}; the full numerical pipeline is summarized in Appendix~\ref{app:numerical_construction}. A momentum-integrated entanglement measure requires an additional weighting by the full two-body momentum distribution and is left for future work.
\begin{figure}[t]
	\centering
	\includegraphics[scale=0.25]{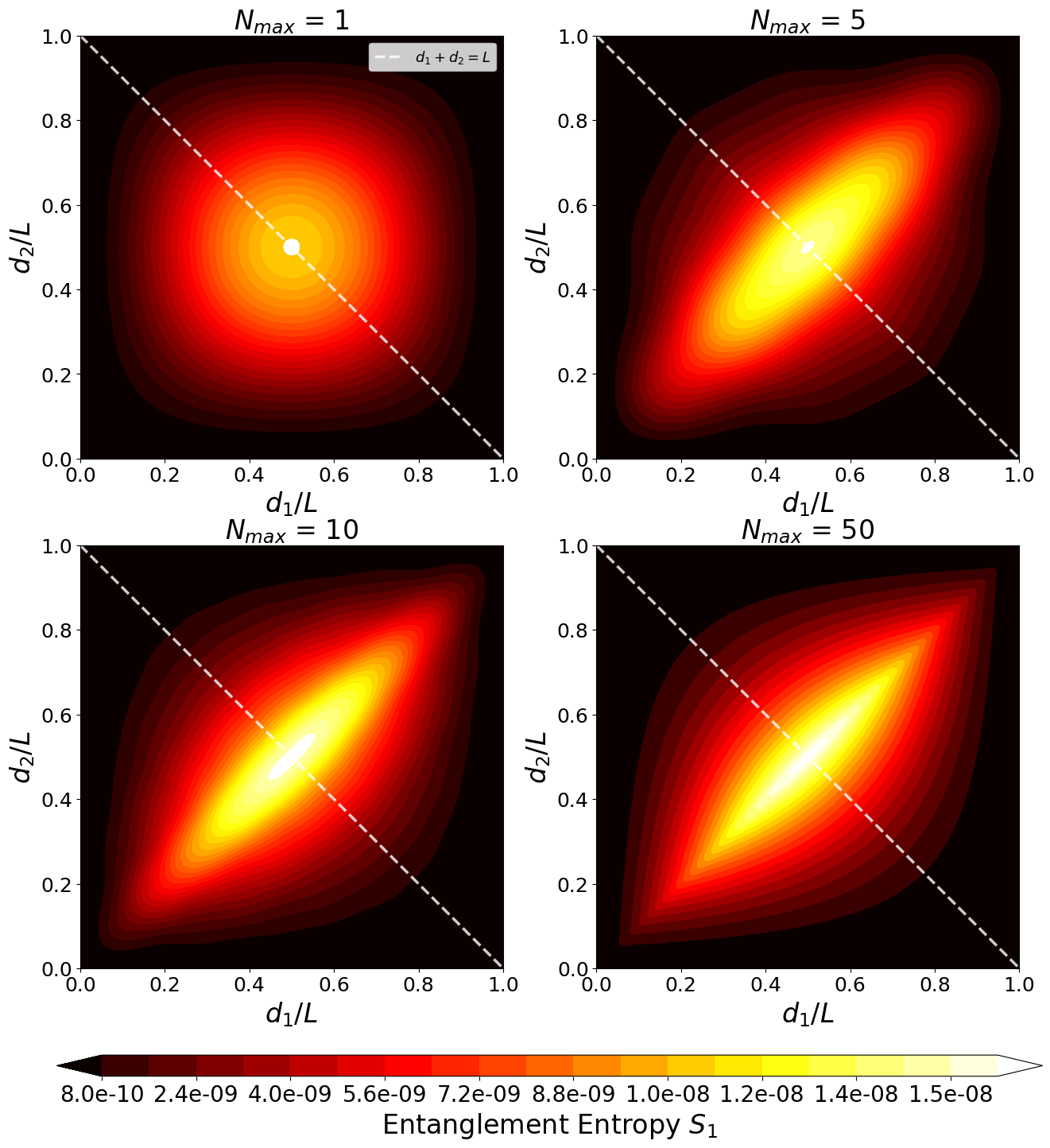}\hfill
	\includegraphics[scale=0.35]{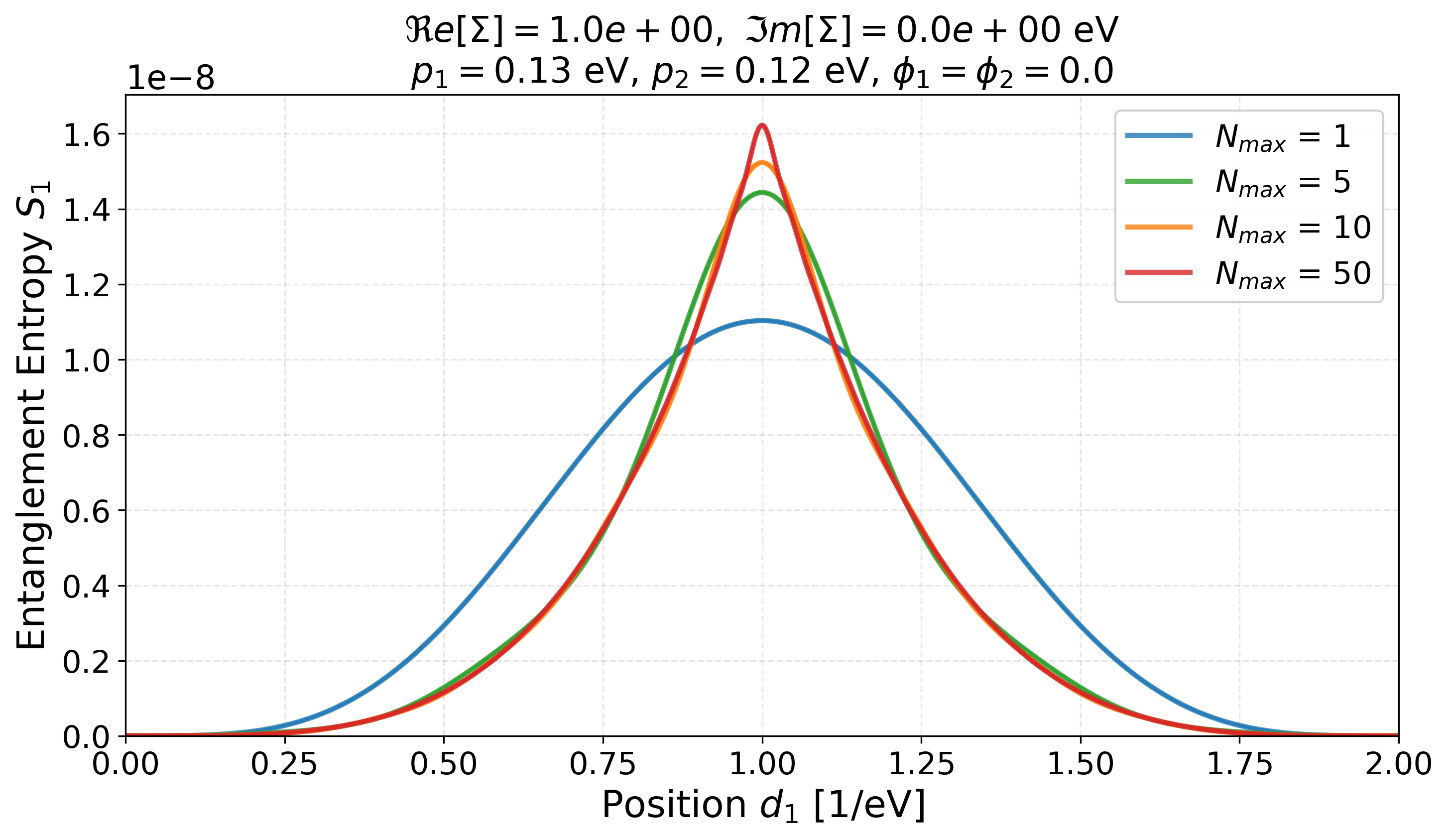}
	\caption{\textit{Top}: Entanglement entropy as a function of the inter-layer distances \(d_1\) and \(d_2\) for different values of the mode cutoff \(N_{\max}\). \textit{Bottom}: one-dimensional cuts along the symmetric configuration \(d_1+d_2=L\). All panels use \(\lambda_{so}=3.9\,\mathrm{meV}\), corresponding to \(m\simeq 2.13\,\mathrm{eV}\) in the numerical convention.}
	\label{fig:entropy_vs_nmax}
\end{figure}
\begin{figure}[ht]
	\centering
	\includegraphics[scale=0.35]{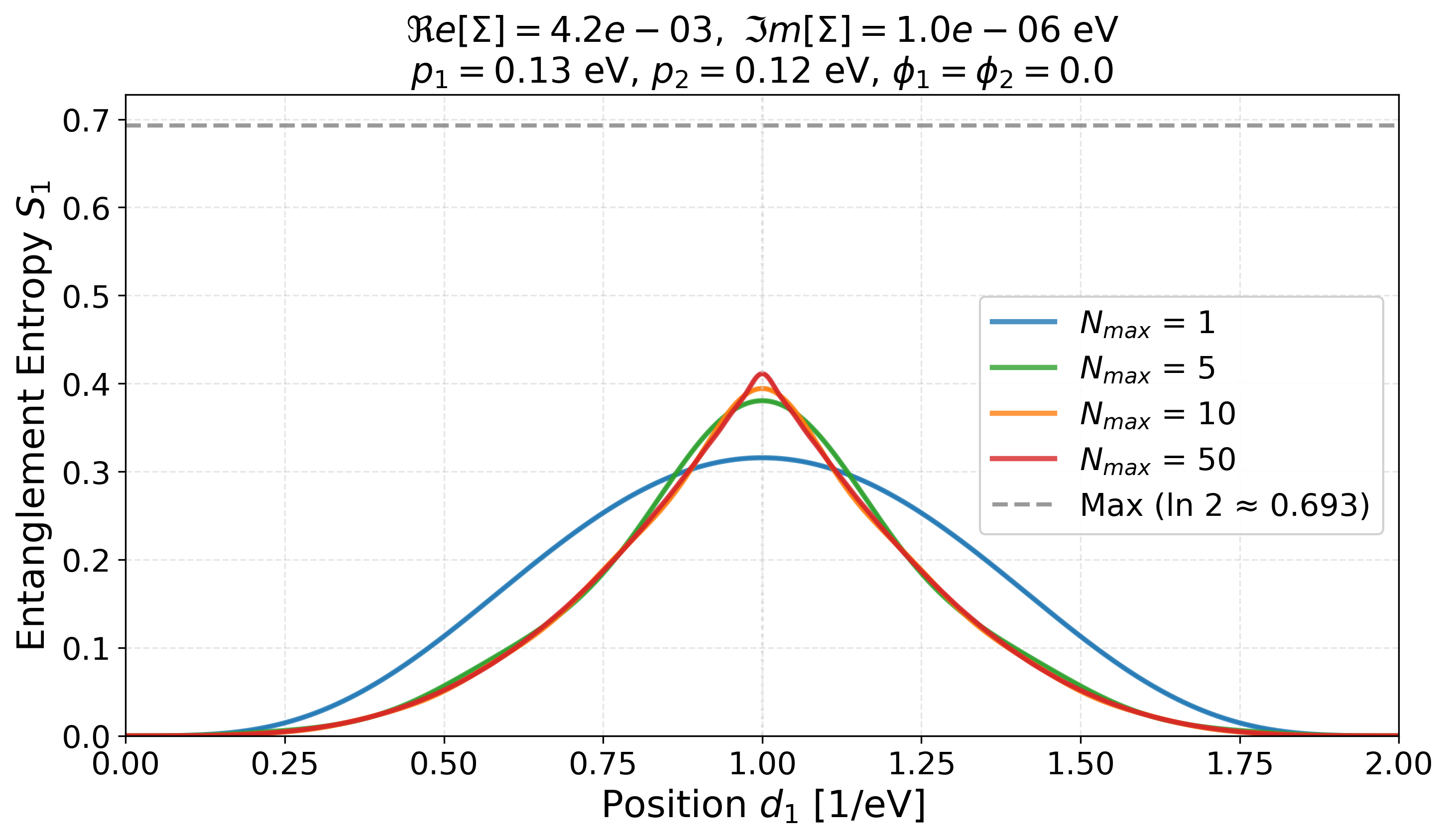}
	\caption{Entanglement entropy $S_1$ as a function of the inter-layer distance $d_1$ along the symmetric configuration $d_1+d_2=L$ for fixed $(\Sigma_1,\Sigma_2)$ and fixed kinematic configuration $(p_1,p_2,\phi_1,\phi_2)$. Here we use \(\Re\Sigma_1=\Re\Sigma_2=4.2\times10^{-3}\,\mathrm{eV}\), chosen from Fig.~\ref{fig:entropy_vs_sigma_log} as a representative value near the entropy maximum while preserving the validity of the Born approximation. All panels also use \(\lambda_{so}=3.9\,\mathrm{meV}\), corresponding to \(m\simeq 2.13\,\mathrm{eV}\) in the numerical convention.}
	\label{fig:entropy_vs_distance_fixed}
\end{figure}

\begin{figure*}[ht]
	\centering
	\includegraphics[width=0.95\textwidth]{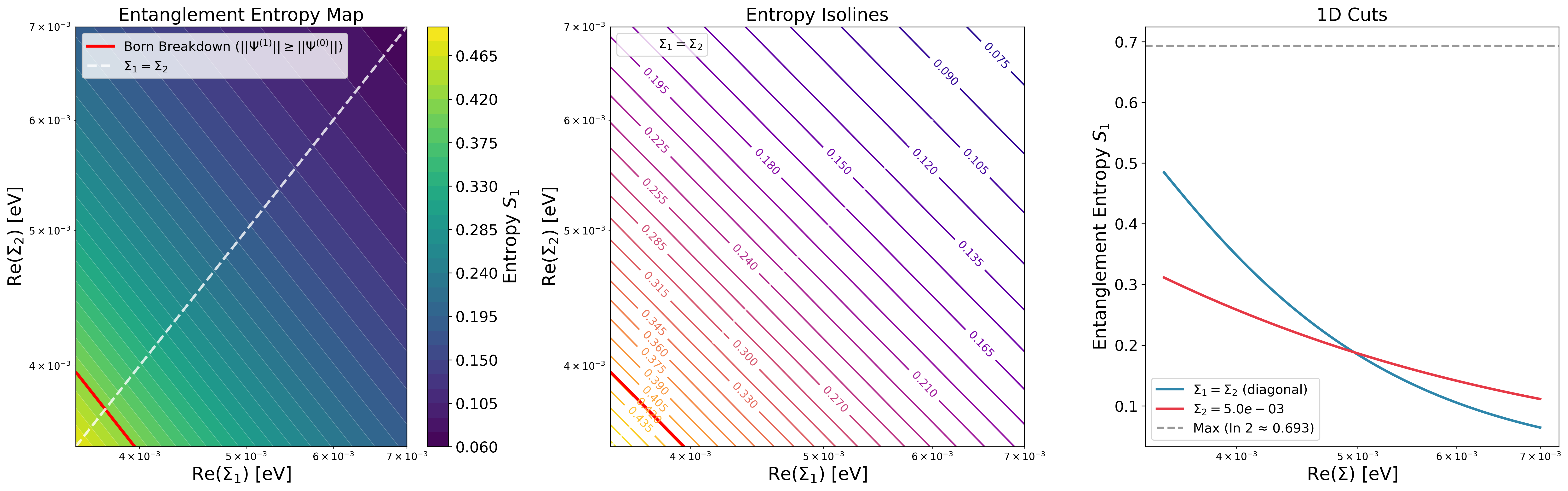}
	\caption{Entanglement entropy $S_1$ as a function of the real parts of the self-energy parameters $\Sigma_1$ and $\Sigma_2$ (logarithmic axes, in eV), for fixed quasiparticle momenta $p_1=0.13$ eV, $p_2=0.12$ eV, collinear angles $\phi_1=\phi_2=0$, and emitter positions $d_1=0.9$, $d_2=1.1$ eV$^{-1}$ inside the cavity. The region to the left of it lies outside the perturbative regime. Results in that region are shown for completeness but should not be interpreted as physical predictions of the Born-level scheme. The dashed white line indicates $\Sigma_1=\Sigma_2$. Parameters: $\lambda_{so}=3.9$ meV ($m\simeq 2.13$ eV in natural units), $\mathrm{Im}(\Sigma)=10^{-6}$ eV.}
	\label{fig:entropy_vs_sigma_log}
\end{figure*}
\section{Results}
\label{sec:results}
The angular coefficients \(\mathbf{I}_1, \;\mathbf{I}_2,\;\mathbf{I}_3\) and \(\mathbf{I}_4\), defined as one-dimensional integrals over the exchanged momentum direction (see Appendix~\ref{app:bs-radial}), were evaluated numerically.
After performing the frequency and radial momentum integrations analytically using Dirac delta constraints, the remaining angular integrals were computed with the Gauss-Legendre quadrature summarized in Appendix~\ref{app:num-angular}.
Typical calculations employed grids of up to \(N_\phi\sim10^3\) quadrature points, ensuring convergence of all wave-function components within numerical precision.

The summation over cavity modes was truncated at a finite cutoff \(N_{\max}\), whose convergence was explicitly tested, as illustrated in Fig.~\ref{fig:entropy_vs_nmax}.
For all other parameter scans shown in the manuscript (self-energy, momenta, and angular dependence), we verified that increasing \(N_\phi\) does not alter the reported entanglement profiles for the momentum grid resolution.

The reduced density matrix \(\rho_1\) is constructed from the numerically evaluated wave-function components using the normalization and spectral-cleaning procedure described in Sec.~\ref{subsec:rhoA} and Appendix~\ref{app:numerical_construction}. 
We have verified that this procedure does not affect the entanglement entropy beyond numerical noise and does not modify any qualitative or quantitative conclusions.

We begin by validating the proposed method through the reproduction of the results obtained in our previous work. As shown in Fig.~\ref{fig:entropy_vs_nmax}, the entanglement entropy as a function of the inter-layer distance exhibits the same behavior reported in Ref.~\cite{facu2}, upon varying the cutoff \(N_{\max}\) in the summation over the normal modes of the microcavity electromagnetic field.

\begin{figure}[htbp]
	\centering
	\includegraphics[width=0.95\linewidth]{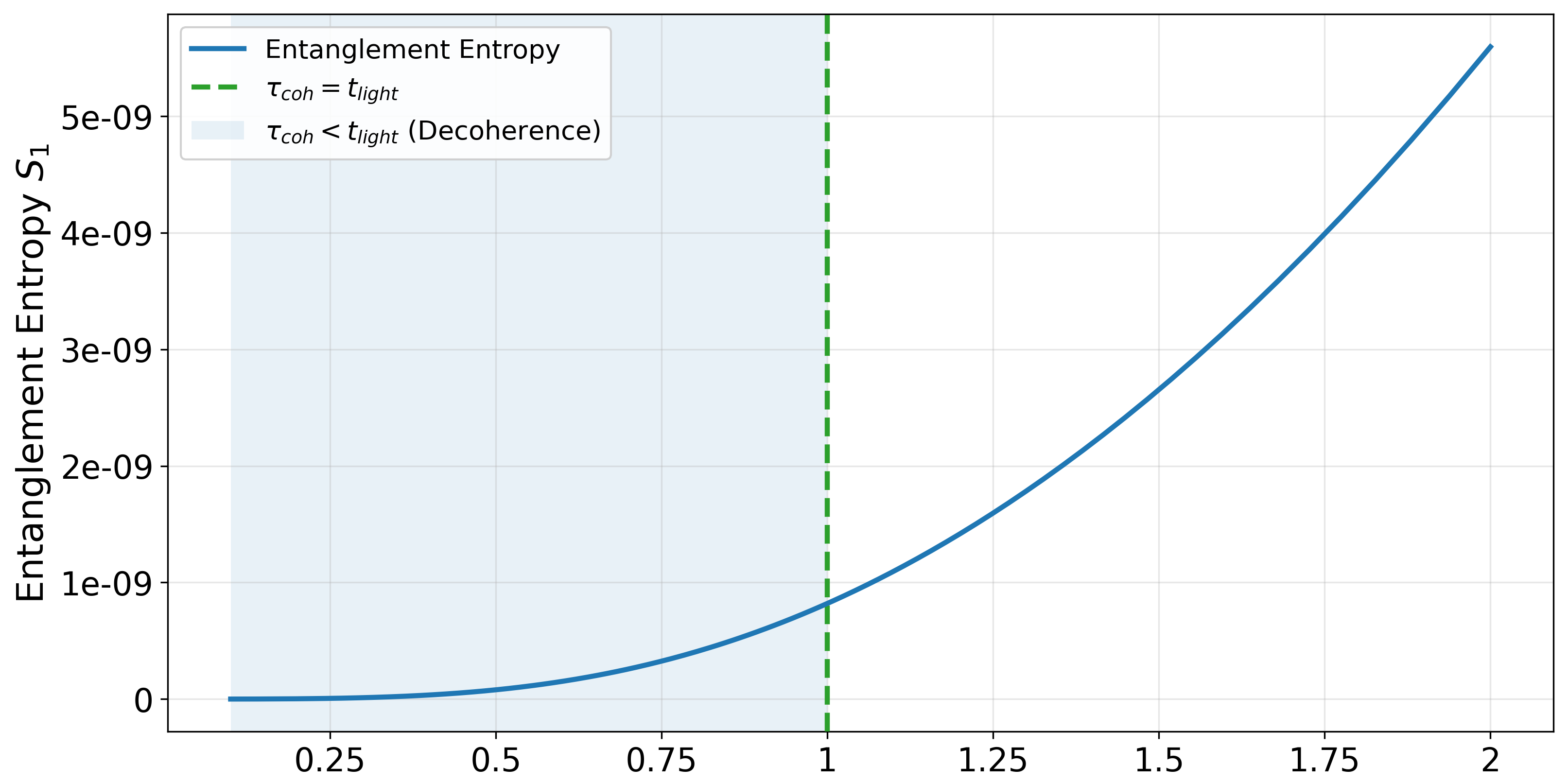}\hfill
	\includegraphics[width=0.95\linewidth]{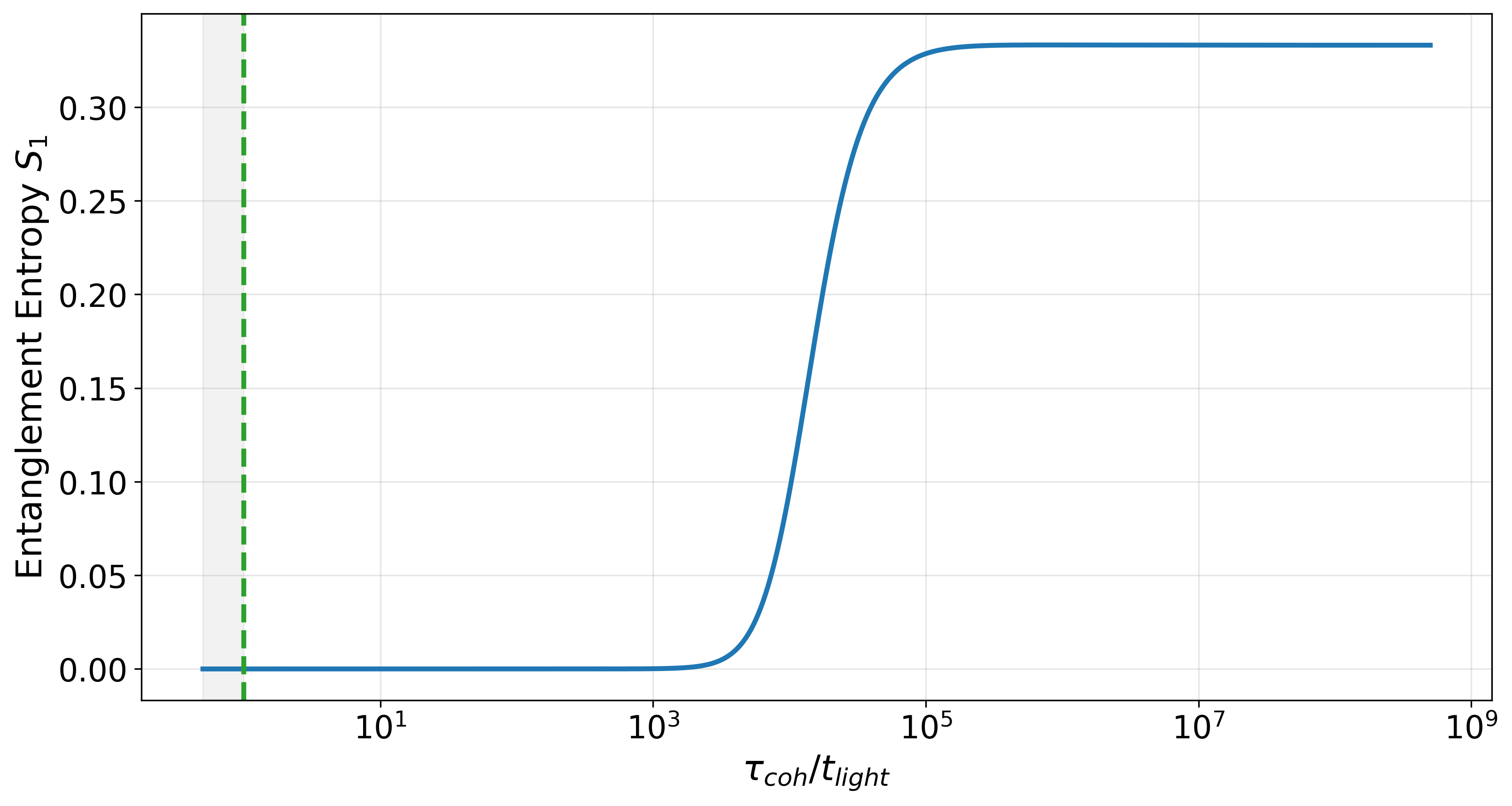}
	\caption{Entanglement entropy as a function of coherence time for $\Re\Sigma$ fixed. \textit{Top}:  a time window close to the light-flight time from one layer to the other. \textit{Bottom}: values where the entanglement entropy stabilizes. Here we fix \(\Re\Sigma_1=\Re\Sigma_2=4.2\times10^{-3}\,\mathrm{eV}\), chosen from Fig.~\ref{fig:entropy_vs_sigma_log} as a representative value near the entropy maximum while preserving the validity of the Born approximation. All panels also use \(\lambda_{so}=3.9\,\mathrm{meV}\).}
	\label{fig:entropy_vs_time}
\end{figure}

For large inter-layer separations, the entanglement entropy becomes strongly suppressed, indicating a weakly correlated regime. However, as shown in Fig.~\ref{fig:entropy_vs_distance_fixed}, this behavior changes drastically once self-energy effects are taken into account. In this regime, modifying the self-energy terms leads to a sudden enhancement of the entanglement entropy.

To characterize this transition in detail, we analyze the entanglement entropy as a function of the electronic self-energies \(\Sigma_1\) and \(\Sigma_2\) of each layer. Fig.~\ref{fig:entropy_vs_sigma_log} presents the entropy landscape on a logarithmic scale, which allows us to capture the behavior over several orders of magnitude. The contour structure reveals a well-defined boundary separating regions of negligible entanglement from a strongly entangled regime. 
We find that when the real part of the self-energies reaches values on the order of \(10^{-3}\,\mathrm{eV}\), the entanglement entropy exhibits an abrupt increase. In addition, a red line is included to mark the regime where the Born approximation ceases to be valid, namely when the real part of the self-energy becomes too small and the first-order perturbative condition \(\|\psi^{(1)}\| \ll \|\psi^{(0)}\|\) is no longer satisfied.

Then, to resolve this feature more clearly, we zoom into the \(\Sigma'\sim10^{-3}\,\mathrm{eV}\) region and analyze the near-threshold coherence-time profile shown in the corresponding results panel by varying the imaginary part of the self-energy while keeping the real part constant. Fig.~\ref{fig:entropy_vs_time} establishes the dissipative window in which the cavity-mediated entanglement predicted by the perturbative treatment survives at long times. For fixed \(\Re\Sigma\), the entanglement entropy becomes stationary only when the quasiparticle coherence time satisfies the strict condition \(\tau_{\mathrm{coh}} > t_{\mathrm{light}}=|d_1 - d_2|/c\). In the strongly dissipative region (left side), coherence is lost before the interaction can entangle the pseudospins. Between these limits, a finite plateau appears, identifying a physically stable regime where cavity QED generates genuine stationary entanglement.
\begin{figure}[htbp]
	\centering
	\includegraphics[width=0.95\linewidth]{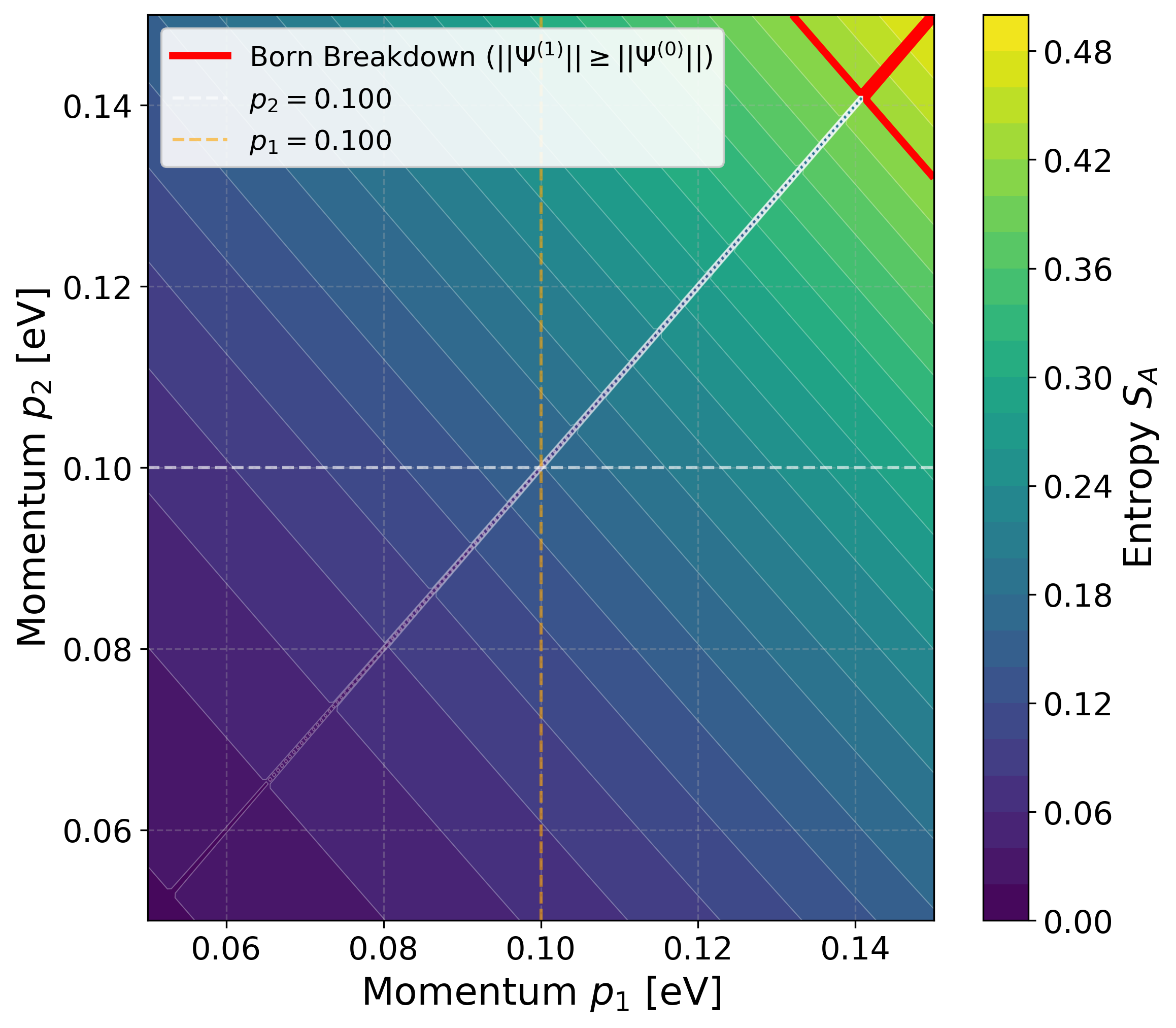}\hfill
	\includegraphics[scale=0.35]{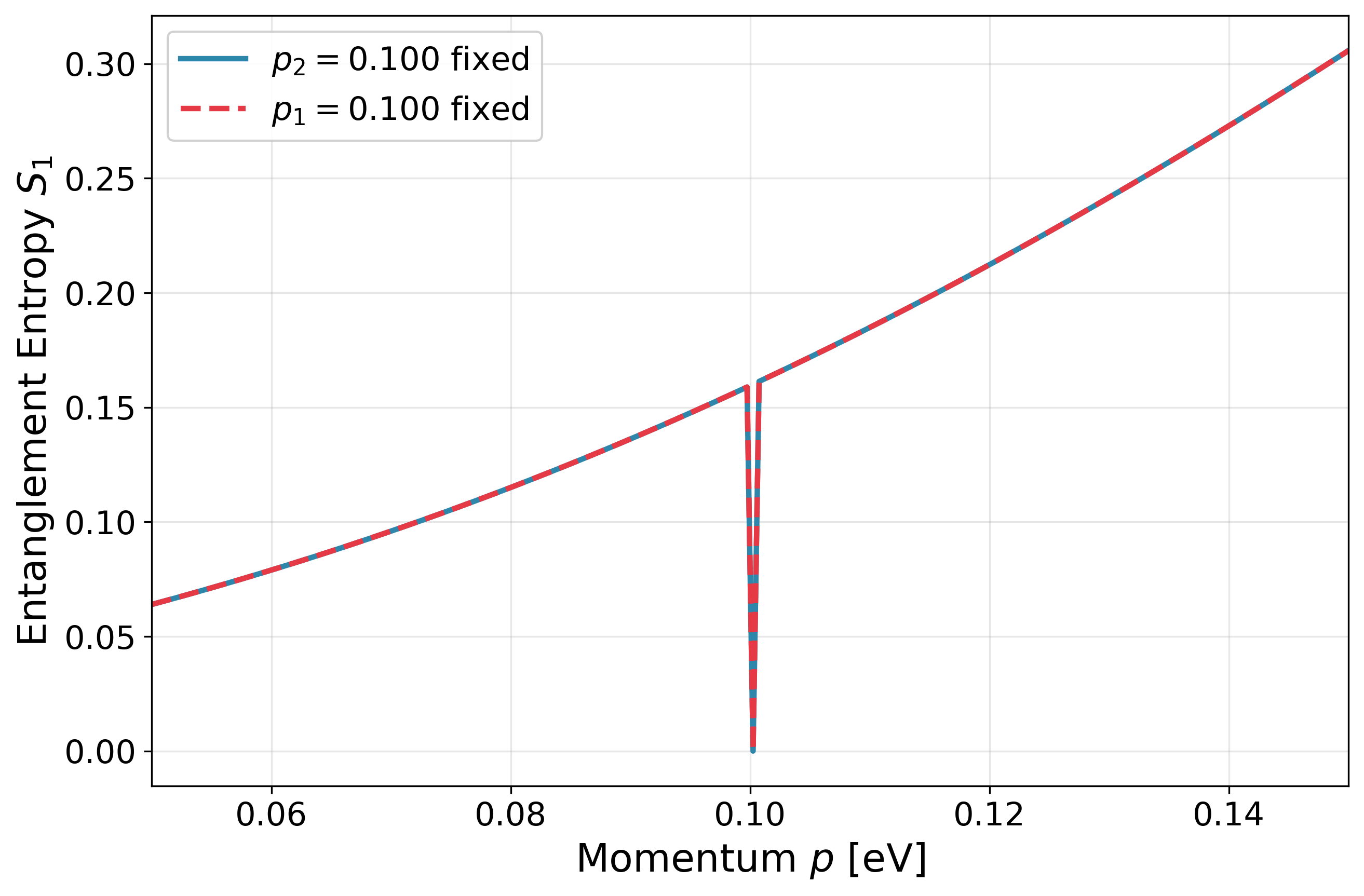}
	\caption{Entanglement entropy $S_1$ as a function of the quasiparticle momenta $p_1$ and $p_2$ (in eV), for fixed collinear angles $\phi_1 = \phi_2 = 0$ and emitter positions $d_1 = 0.9,\ d_2 = 1.1\ \mathrm{eV}^{-1}$ inside the cavity. \textit{Top}: full 2D map; the red contour marks the Born validity boundary ($\|\Psi^{(1)}\| = \|\Psi^{(0)}\|$), beyond which results lie outside the perturbative regime. \textit{Bottom}: 1D cuts at fixed $p_1 = p^{\mathrm{mid}}$ and $p_2 = p^{\mathrm{mid}}$, where $p^{\mathrm{mid}} = 0.10$ eV is the midpoint of the scanned range. Parameters: $\lambda_{so} = 3.9$ meV, $\mathrm{Re}(\Sigma) = 4.2$ meV, $\mathrm{Im}(\Sigma) = 10^{-6}$ eV.}
	\label{fig:entropy_vs_momentum}
\end{figure}

As a final result, we investigate the dependence of the entanglement entropy on the electronic momenta of the two layers. Fig.~\ref{fig:entropy_vs_momentum} shows that the momentum dependence introduces an additional modulation of the entanglement structure, while preserving the sharp transition observed in the self-energy parameter space. This indicates that the enhancement of entanglement is a robust feature, not restricted to a fine-tuned set of dynamical parameters. A distinctive feature of Fig.~\ref{fig:entropy_vs_momentum} is the pronounced dip along the diagonal $p_1=p_2$. In this configuration, the two quasiparticles in the double-layer system are traveling parallel to each other. For the cavity to mediate an interaction, the quasiparticles must exchange a virtual photon, which fundamentally requires a transfer of momentum and energy between the two layers. However, when the particles are perfectly synchronized in their kinematic states, on-shell energy-momentum conservation strictly forbids this exchange—there is zero available phase space to emit and absorb a photon. Because no virtual photon can be exchanged, the quasiparticles do not notice each other through the cavity field and propagate as entirely independent, non-interacting free particles. Without this interaction to correlate their degrees of freedom, the two-body quantum state remains perfectly separable, resulting in an exact zero for the entanglement entropy. Away from this diagonal, there is a finite momentum transfer, and the entanglement is rapidly restored.

\section{Discussion}\label{sec:discussion}
The results presented above reveal a remarkable structure of momentum-space entanglement emerging from the interplay between the microcavity field, the quasiparticle self-energy, and the spinor geometry of two-dimensional Dirac materials. A central numerical observation is a sharp rise in the first-order entropy as the phenomenological self-energy is reduced. Within the first-order scheme, this behavior should be interpreted cautiously: it reflects the proximity of quasiparticle denominators to poles in the linear system used to compute $\Psi^{(1)}$, which amplifies selected spinor components and strongly modifies the normalized conditional state. This amplification is not, by itself, the signature of a physical phase transition; rather, it indicates a regime where the single-iteration truncation becomes delicate and where higher-order or resummed treatments may be required.

It is useful to place these findings in the broader context of perturbative entanglement-harvesting studies. In standard detector-based protocols and in previous vacuum-mediated analyses of honeycomb materials, the harvested correlations in the weak-coupling regime are intrinsically small, providing a numerical baseline set by virtual-particle exchange in the absence of resonant mechanisms~\cite{perche,facu2}. Our present methodology is not identical to those constructions: here we compute a momentum-resolved von Neumann entropy from a normalized conditional two-body state obtained from the first iterative correction to the Bethe--Salpeter amplitude, rather than a detector negativity from a second-order expansion of the reduced density matrix. Nevertheless, when quasiparticle dressing is absent or negligible, our results remain in the same perturbative low-entanglement regime, as is already visible by comparing Figs.~\ref{fig:entropy_vs_nmax} and~\ref{fig:entropy_vs_distance_fixed}: the former reproduces the weak-entanglement baseline of the cavity-mediated problem, whereas the latter shows that, once self-energy effects are taken into account, entropy increases significantly. The crossover is quantified more directly in Fig.~\ref{fig:entropy_vs_sigma_log}, where the self-energy scan resolves the transition in parameter space and reveals that the phenomenological self-energy acts as the tuning parameter that drives the system out of the weak-coupling baseline as the dressed quasiparticle denominators move toward resonance.

Importantly, this crossover is not limited to a fine-tuned choice of dynamical variables. As shown in Section~\ref{sec:results}, the enhancement of entanglement persists over a broad range of electronic momenta, indicating that the effect is robust against variations in cavity-mediated interaction. This robustness suggests that the phenomenon could be experimentally accessible, provided that the cavity geometry and material parameters allow for sufficient control of self-energy effects.

Taking $\Sigma$ to be purely real corresponds to the ideal limit of infinite quasiparticle lifetime. In a more realistic dissipative description with a complex self-energy $\Sigma \to \Sigma' + i\Gamma$, a finite imaginary part $\Gamma$ would regularize the poles and broaden the sharp crossover into a finite-width resonance peak. The causal condition identified in Section~\ref{sec:results}---namely that stationary entanglement requires the quasiparticle coherence time $\tau_{\mathrm{coh}} = 1/\Gamma$ to exceed the photon travel time between the layers---places a concrete lower bound on the dissipation that a physical implementation can tolerate. The results reported here should therefore be interpreted as the idealized resonance limit of a physical enhancement that, under experimental conditions, would appear as a broadened but observable feature.

Beyond the role of self-energy renormalization, our results highlight the influence of geometric constraints on entanglement generation. The strongest suppression occurs when both quasiparticles have nearly identical propagation directions and magnitudes, for which the dominant exchanged momentum is small, and spinor mixing is reduced. The physical origin of this suppression is clear. When the two quasiparticles share identical kinematic states, there is no available phase space for the cavity to mediate a momentum transfer: on-shell energy-momentum conservation strictly forbids the emission and reabsorption of a virtual photon, and without that exchange the two layers decouple entirely, leaving the two-body state perfectly separable along the diagonal $p_1 = p_2$. For antialigned configurations, the momentum transfer is finite, and the suppression mechanism is qualitatively different, as reflected in the asymmetric structure of Fig.~\ref{fig:entropy_vs_momentum}.

From a broader perspective, our findings demonstrate how concepts traditionally associated with relativistic QFT—self--energy renormalization, virtual particle exchange, and spinor geometry—manifest themselves in condensed matter systems under cavity QED conditions, and how they can be turned into a quantitative diagnostic of quantum-information resources. More specifically, the emergence of entropy values close to the maximum allowed by the chosen pseudospin bipartition suggests that cavity-dressed quasiparticle states can approach Bell-like correlations in the sublattice sector. Although our first-iteration treatment does not yet solve the homogeneous Bethe-Salpeter equation and therefore cannot by itself identify stationary entangled eigenstates, it pinpoints the region of parameter space where a non-perturbative extension should look for such states. The framework is not limited to the ladder approximation and single-photon exchange considered here: it can be extended to higher-order processes, different cavity geometries, and alternative interaction channels, offering a systematic roadmap toward engineering Bell-type resources in Dirac materials for quantum-information applications.

\section{Conclusions}\label{sec:conclusions}

We have investigated the momentum-space entanglement in a two-body Dirac-quasiparticle state in a double-layer honeycomb lattice embedded in a planar electromagnetic cavity. Modeling the low-energy excitations as massive Dirac fermions and employing the Bethe-Salpeter equation within a cavity QED framework, we computed the reduced density matrix and the associated conditional pseudospin von Neumann entropy as a quantitative measure of entanglement. We found that this quantity exhibits a crossover from the perturbative low-entanglement baseline—consistent with the small values reported in vacuum-mediated harvesting protocols~\cite{pozas,perche} and in previous cavity-QED analyses of honeycomb lattices~\cite{facu1,facu2}—to a strongly enhanced regime as the phenomenological self-energy is tuned, with the largest values approaching the maximum allowed by the chosen pseudospin bipartition; low-entropy regions remain associated with near-identical quasiparticle momenta.

We further found that stationary entanglement requires the quasiparticle coherence time to exceed the photon propagation time between the two layers of the cavity. This causal condition is consistent with graphene-based teleportation proposals, where coherent state transfer across remote nodes likewise requires the communication time window to remain shorter than the decoherence time~\cite{asjad2021teleport,asjad2022network}, and it suggests that decoherence-robust indicators such as uncertainty-induced non-locality~\cite{mohamed2022nonlocal} may serve as useful complements to entropic diagnostics when characterizing the dissipative regimes identified here.

These findings provide a connection between modelling approaches inspired by relativistic QFT and the analysis of quantum information in Dirac materials, while clarifying the limitations of this methodology. A fully predictive framework should incorporate self-consistent Bethe-Salpeter/Dyson dressing, complex and momentum-dependent self-energies, and momentum-integrated entanglement measures. Most importantly, solving the homogeneous Bethe-Salpeter equation is necessary to determine whether the near-maximal-entropy regime identified here corresponds to genuine Bell-type quasiparticle states or merely represents a precursor to strong entanglement. The current formalism provides a concrete roadmap for this research direction by isolating the cavity geometry, momentum, and dressing conditions under which Dirac materials may be driven toward the highly entangled resources relevant for quantum-information technologies. In doing so, it establishes a systematic approach for engineering Bell-type states in solid-state platforms.

\begin{acknowledgments}
We thank Hugo Hernandez and Gabriel Jalil for discussions.
This article was partially supported by CONICET grants (Argentina National Research Council) and Universidad Nacional del Sur (UNS). F.A. J.S.A. and A. G. acknowledge support as members of CONICET and Departamento de Física, Universidad Nacional del Sur. 
\end{acknowledgments}

\bibliography{bib-refs}

\onecolumngrid
\appendix
\setcounter{secnumdepth}{2}
\renewcommand{\thesubsection}{\thesection\arabic{subsection}}
\section{Bethe--Salpeter Kernel Derivation}
\label{app:bethe_salpeter}

\noindent
\textit{In Appendix~\ref{app:bs-kernel}, we derive the effective Bethe-Salpeter kernel in momentum space; in Appendices~\ref{app:bs-frequency} and~\ref{app:bs-radial}, we perform the frequency and radial momentum integrations; and in Appendix~\ref{app:bs-spinor}, we obtain the explicit spinor structure employed in the calculation of the reduced density matrix.}

\subsection{Momentum-Space Representation of the Kernel}
\label{app:bs-kernel}

The Fourier transform of the Bethe-Salpeter kernel is defined as
\begin{align}
\bar{K}^{12}(p'_1,p'_2;p_1,p_2)
&=
\frac{1}{(2\pi)^6}
\int d^3 r'_1 d^3 r'_2 d^3 r_1 d^3 r_2\, e^{i(p'_1 r'_1 + p'_2 r'_2 - p_1 r_1 - p_2 r_2)}
K^{12}(r'_1,r'_2;r_1,r_2).
\end{align}

Substituting the real-space kernel and retaining the single-photon exchange contribution, we obtain
\begin{align}
\bar{K}^{12} (p'_1,p'_2;p_1,p_2)
&\approx
\frac{(-ie)^2}{(2\pi)^6}
\int d^3 r_1 d^3 r_2\, e^{i[r_1(p'_1-p_1)+r_2(p'_2-p_2)]}
\gamma_\mu^a
D_F^{\mu\nu}(r_1,r_2)
\gamma_\nu^b .
\end{align}

Using the cavity photon propagator and defining
\begin{equation}
\zeta_n \equiv
\zeta^2
\sin\!\left(\frac{n\pi d_1}{L}\right)
\sin\!\left(\frac{n\pi d_2}{L}\right),
\end{equation}
the kernel in momentum space becomes
\begin{equation}
\bar{K}^{12} (p'_1,p'_2;p_1,p_2)
=
(-ie)^2
\int \frac{d^3 q}{(2\pi)^3}
\mathcal{D}(q)\,
\Gamma^{12}\,
\delta^{(3)}(p'_1-p_1-q)
\delta^{(3)}(p'_2-p_2+q),
\end{equation}
where we introduced the effective cavity propagator
\begin{equation}
\mathcal{D}(q)
\equiv
\sum_{n=0}^{\infty}
\frac{\zeta_n}{q^2-(n\pi/L)^2+i\epsilon},
\end{equation}
and the effective Dirac vertex
\begin{equation}
\Gamma^{ij}
\equiv
\gamma_\mu^i g^{\mu\nu} \gamma_\nu^j
=
\gamma_0^i\gamma_0^j
-\gamma_1^i\gamma_1^j
-\gamma_2^i\gamma_2^j .
\end{equation}

\subsection{Two-Particle Wave Function in Momentum Space}
\label{app:bs-wavefunction}

The free two-body wave function in real space reads
\begin{align}
\psi^0_{12}(r_1,r_2)
&=
\int \frac{d^2 p'_1 d^2 p'_2}{(2\pi)^4} 
\Big[
u_1(\vb{p}'_1)e^{-ip'_1 r_1}
+
v_1(\vb{p}'_1)e^{ip'_1 r_1}
\Big] \otimes
\Big[
u_2 (\vb{p'}_2)e^{-ip'_2 r_2}
+
v_2 (\vb{p}'_2)e^{ip'_2 r_2}
\Big].
\end{align}

Fourier transforming, one obtains
\begin{align}
\varphi^0_{12}(p_1,p_2)
&=
u_1(\vb{p}_1)
\delta\!\left(p_1^0-\frac{E_{\vb{p}_1}}{v_{F1}}\right)
\otimes
u_2(\vb{p}_2)
\delta\!\left(p_2^0-\frac{E_{\vb{p}_2}}{v_{F2}}\right)
+\cdots .
\end{align}

For compactness, we now display only the conduction-band electron--electron (\textit{ee}) sector, in which both layers carry positive-energy spinors $u_i$ and on-shell energies $p_i^0=+E_i/v_{Fi}$. The electron--hole, hole--electron, and hole--hole sectors are obtained by replacing one or both electron spinors by the corresponding hole spinors $v_i$, and all four sectors are included in the numerical construction of the total state. The interacting wave function entering the Bethe--Salpeter equation then reads
\begin{align}
\varphi_{12}(p_1-q,p_2+q)
&=
u_1 (\vb{p}_1-\vb{q})
\otimes
u_2(\vb{p}_2+\vb{q})
\delta\!\left(p_1^0-q^0-\frac{E_{\vb{p}_1-\vb{q}}}{v_{F1}}\right)
\delta\!\left(p_2^0+q^0-\frac{E_{\vb{p}_2+\vb{q}}}{v_{F2}}\right).
\end{align}

\subsection{Frequency Integration}
\label{app:bs-frequency}

At the level of the two-particle amplitude used in this work, the intermediate quasiparticle lines are treated in a quasiparticle (on-shell) approximation. This is reflected by the Dirac delta functions enforcing the on-shell dispersion for each constituent, which allows the \(q^0\) integration to be carried out explicitly. Integrating over the frequency component \(q^0\), we obtain
\begin{align}
\int \frac{d^2 q}{(2\pi)^2}
\mathcal{D}\left(p_1^0-\frac{E_{\vb{p}_1 - \vb{q}}}{v_{F1}}\right)\,
\Gamma^{12}
[u_1(\vb{p}_1-\vb{q}) \otimes u_2(\vb{p}_2+\vb{q})]\,
\delta\!\left(\mathcal{F}(q)\right),
\end{align}
where we defined the kinematic constraint
\begin{equation}
\mathcal{F}(q)
\equiv
p_1^0+p_2^0
-\frac{E_{\vb{p}_1-\vb{q}}}{v_{F1}}
-\frac{E_{\vb{p}_2+\vb{q}}}{v_{F2}}.
\end{equation}
In deriving this form, the two original on-shell delta functions are combined by integrating over $q^0$, yielding a single constraint $\delta(\mathcal{F}(q))$ for the radial momentum variable.

\subsection{Angular and Radial Momentum Integration}
\label{app:bs-radial}

Using the identity
\begin{equation}
\delta(f(x))
=
\sum_{x_0}
\frac{\delta(x-x_0)}{|f'(x_0)|},
\end{equation}
and writing $d^2 q = q\,dq\,d\phi_{\vb q}$, the constraint $\mathcal{F}(q)=0$ yields two roots for the radial variable. One of them is $q_0=0$; its contribution vanishes because it is multiplied by the radial Jacobian $q$. Therefore, only the non-trivial root contributes to the radial integral. Defining $E_i \equiv \sqrt{(v_{Fi}p_i)^2+\lambda_i^2}$, the surviving root can be written as
\begin{equation}
q_0^{(\pm)}
=
\frac{
2\left(E_1\pm E_2\right)
\left(
E_2\,p_1\cos(\phi_1-\phi_{\vb q})
\mp
E_1\,p_2\cos(\phi_2-\phi_{\vb q})
\right)
}{
\left(E_1\pm E_2\right)^2
-
\left(
 p_1\cos(\phi_1-\phi_{\vb q}) + p_2\cos(\phi_2-\phi_{\vb q})
\right)^2
},
\end{equation}
where the sign choice depends on the channel under consideration (+ to $ee$, $hh$ and $-$ to $eh$, $he$). For notational simplicity, we denote this surviving root again by $q_0$ in the final integral. The radial momentum integration can then be carried out, yielding
\begin{equation}\label{the_int}
\int_0^{2\pi}
\frac{d\phi_{\vb{q}}}{(2\pi)^2}
\frac{\mathcal{D}\left(p_1^0-\frac{E_{\vb{p}_1 - q_0 \hat{\vb{q}}}}{v_{F1}}\right)}{|\mathcal{F}'(q_0)|}
\Gamma^{12}
\big[
u_1(\vb{p}_1-q_0\hat{\vb{q}})
\otimes
u_2(\vb{p}_2+q_0\hat{\vb{q}})
\big].
\end{equation}

\subsection{Spinor Structure}
\label{app:bs-spinor}

Defining  $\phi_{1q}\equiv \phi_{\vb{p}_1-\vb{q}}$ and $\phi_{2q}\equiv \phi_{\vb{p}_2+\vb{q}}$, and still restricting to the displayed \textit{ee} sector, the action of the effective vertex on the spinor product yields
\begin{equation}
\Gamma^{12}
\left[
u_1(\vb{p}_1-\vb{q})
\otimes
u_2(\vb{p}_2+\vb{q})
\right]
=
\frac{i}{\mathcal{N}}
\begin{pmatrix}
-\chi_1^{+}\chi_2^{+} e^{i(\phi_{1q}+\phi_{2q})} \\
i\chi_1^{+} e^{i\phi_{1q}} \\
i\chi_2^{+} e^{i\phi_{2q}} \\
1
\end{pmatrix},
\end{equation}
with normalization
\begin{equation}
\mathcal{N}
=
\sqrt{\left[1+\left(\chi_1^{+}\right)^2\right]
\left[1+\left(\chi_2^{+}\right)^2\right]},
\end{equation}
and
\begin{align}
\chi_1^{\pm}(\vb{p}_1-\vb{q})
&=
\frac{v_{F1}|\vb{p}_1-\vb{q}|}
{\pm E(\vb{p}_1-\vb{q})+\lambda_1},\\
\chi_2^{\pm}(\vb{p}_2+\vb{q})
&=
\frac{v_{F2}|\vb{p}_2+\vb{q}|}
{\pm E(\vb{p}_2+\vb{q})+\lambda_2}.
\end{align}
The remaining channels entering the numerics are obtained from the same algebraic structure by the replacements $\left(u_1,u_2\right)\to \left(u_1,v_2\right)$, $\left(v_1,u_2\right)$, and $\left(v_1,v_2\right)$ for the $eh$, $he$, and $hh$ sectors, respectively. Equivalently, one replaces $(\chi_1^{+},\chi_2^{+})\to (\chi_1^{+},\chi_2^{-})$, $(\chi_1^{-},\chi_2^{+})$, and $(\chi_1^{-},\chi_2^{-})$, together with the corresponding on-shell choices $p_i^0=\pm E_i/v_{Fi}$ in each layer.
\section{Numerical Construction of the Two-Body State and Entropy Calculation}\label{app:numerical_construction}

\textit{Throughout the numerical sections, we adopt the following
rescaled variables (natural units $\hbar = c = 1$, eV--nm system):
$m_i \equiv \lambda_i / v_{Fi}$ denotes the Dirac mass of layer $i$. Since spin is not resolved in the present treatment, explicit spin labels are suppressed throughout this appendix; any spin dependence is understood to be absorbed into the effective parameter $m_i$.}
\subsection{Numerical Evaluation of the Angular Integrals}
\label{app:num-angular}

After discretizing Eq. \eqref{the_int}, the angular integrals are computed using Gauss-Legendre quadrature. 
For an integral of the form
\begin{equation}
I_\alpha = \int_{0}^{2\pi} d\phi \, \mathcal{F}_\alpha(\phi;\mathbf{p}),
\end{equation}
we map the Gauss-Legendre nodes $x_i \in [-1,1]$ to the physical interval $[0,2\pi]$ through
\begin{equation}
\phi_i = \pi(x_i+1),
\qquad
w_i^{(\phi)} = \pi w_i ,
\end{equation}
where $w_i$ are the standard Legendre weights. The integral is then approximated as
\begin{equation}
I_\alpha \approx \sum_{i=1}^{N_\phi}
w_i^{(\phi)}\, \mathcal{F}_\alpha(\phi_i;\mathbf{p}),
\end{equation}
with $N_\phi$ quadrature points.
For each interaction channel 
\[
\mathcal{C} \in \{ee,eh,he,hh\},
\]
four components $(I_1,I_2,I_3,I_4)$ are computed, corresponding to the four components of the two-body spinor structure.

\subsection{Unperturbed Two-Body Wave Function}
\label{app:num-free-state}

Each layer $i=1,2$ is described by a two-component spinor. 
For a given momentum $p_i$, effective mass $m_i \equiv \lambda_i / v_{Fi}$ and polar angle $\phi_i$, we define
\begin{equation}
\chi_{i}^{\pm}(p_{i})
=
\frac{p_i}{\pm \sqrt{p_i^2+m_{i}^2}+m_{i}}.
\end{equation}
The corresponding normalized spinors read
\begin{equation}
u_i(p_{i},\phi_{i})
=
\frac{1}{\sqrt{1+\left(\chi_{i}^{+}\right)^2}}
\begin{pmatrix}
1\\
\chi_{i}^{+}\, e^{i\phi_{i}}
\end{pmatrix},
\qquad
v_i(p_{i},\phi_{i})
=
\frac{1}{\sqrt{1+\left(\chi_{i}^{-}\right)^2}}
\begin{pmatrix}
1\\
\chi_{i}^{-}\, e^{i\phi_{i}}
\end{pmatrix}.
\end{equation}
The spinor $u_i$ corresponds to an electron excitation, while $v_i$ corresponds to a hole excitation. It is convenient to label the four channels by $ee$, $eh$, $he$, and $hh$, according to whether each layer carries a $u$ or $v$ spinor. For completeness, the unperturbed contribution entering the coherent superposition can be written as
\begin{equation}
\begin{aligned}
\Psi^{(0)}_{\text{tot}}=&\sum_{\mathcal{C}} 
\Psi^{(0)}_{\mathcal{C}}=\Psi^{(0)}_{ee}
+\Psi^{(0)}_{eh}
+\Psi^{(0)}_{he}
+\Psi^{(0)}_{hh}\\
=&\quad u_1(p_1,\phi_1)
\otimes
u_2(p_2,\phi_2)
+
u_1(p_1,\phi_1)
\otimes
v_2(p_2,\phi_2)\\
&+
v_1(p_1,\phi_1)
\otimes
u_2(p_2,\phi_2)
+
v_1(p_1,\phi_1)
\otimes
v_2(p_2,\phi_2).
\end{aligned}
\end{equation}
Explicitly, in the product basis,
\begin{equation}
\Psi^{(0)}_{\text{tot}}=
\begin{pmatrix}
u_{1A} \, u_{2A} \\
u_{1A} \, u_{2B} \\
u_{1B} \, u_{2A} \\
u_{1B} \, u_{2B}
\end{pmatrix}
+
\begin{pmatrix}
u_{1A} \, v_{2A} \\
u_{1A} \, v_{2B} \\
u_{1B} \, v_{2A} \\
u_{1B} \, v_{2B}
\end{pmatrix}
+
\begin{pmatrix}
v_{1A} \, u_{2A} \\
v_{1A} \, u_{2B} \\
v_{1B} \, u_{2A} \\
v_{1B} \, u_{2B}
\end{pmatrix}
+
\begin{pmatrix}
v_{1A} \, v_{2A} \\
v_{1A} \, v_{2B} \\
v_{1B} \, v_{2A} \\
v_{1B} \, v_{2B}
\end{pmatrix}.
\end{equation}

\subsection{First-Order Correction from the Bethe--Salpeter Equation}
\label{app:num-first-order}

The correction $\Psi^{(1)}$ is obtained by solving a linear system of the form
\begin{equation}
G^{-1}_{\text{tot},\mathcal{C}} \, \Psi^{(1)}_{\mathcal{C}}
=
I_{\mathcal{C}},
\end{equation}
where $\mathcal{C}\in\{ee,eh,he,hh\}$ and $I_{\mathcal{C}}$ is the vector formed by the four numerically evaluated integrals, ordered as
\begin{equation}
\Psi^{(1)}_{\mathcal{C}} =
\begin{pmatrix}
\varphi_{AA} \\
\varphi_{AB} \\
\varphi_{BA} \\
\varphi_{BB}
\end{pmatrix},
\qquad
I_{\mathcal{C}}=
\begin{pmatrix}
I_{AA} \\
I_{AB} \\
I_{BA} \\
I_{BB}
\end{pmatrix}.
\end{equation}
Then it is convenient to write the left/right action completely in indicial notation, without introducing any explicit vectorization operation:
\begin{equation}
\left(S_1^{-1}\,\varphi\,S_2^{-1}\right)_{\alpha\beta}
=
\left(S_1^{-1}\right)_{\alpha\gamma}\,\varphi_{\gamma\delta}\,\left(S_2^{-1}\right)_{\delta\beta}.
\end{equation}
Using
\begin{equation}
\left(S_2^{-1}\right)_{\delta\beta}=
\left[\left(S_2^{-1}\right)^{T}\right]_{\beta\delta},
\end{equation}
the corresponding operator acting on the composite index space is precisely
\begin{equation}
G^{-1}_{\text{tot},\mathcal{C}}
=
S_1^{-1} \otimes (S_2^{-1})^{T},
\end{equation}
which is a \(4\times 4\) matrix. The single-layer inverse propagators are written as
\begin{equation}
S_i^{-1} =
\begin{pmatrix}
p^{0}_{i} - m_i -\Sigma_i & - p_{i} e^{-i\phi_i} \\
p_{i}e^{i\phi_i} & -p^{0}_{i} -m_i - \Sigma_i
\end{pmatrix},
\end{equation}
with on-shell frequency choice $p_i^0=+E_i/v_{Fi}$ when layer $i$ carries $u_i$ and $p_i^0=-E_i/v_{Fi}$ when it carries $v_i$ in the channel $\mathcal{C}$. The correction is therefore obtained by
\begin{equation}
\Psi^{(1)}_{\mathcal{C}}
=
\left(G^{-1}_{\text{tot},\mathcal{C}}\right)^{-1}
I_{\mathcal{C}}.
\end{equation}

\subsection{Coherent Superposition of Interaction Channels}
\label{app:num-superposition}

The total state is constructed as a coherent superposition over the four interaction channels,
\begin{equation}
\Psi_{\text{tot}}
=
\sum_{\mathcal{C}} 
w_{\mathcal{C}}
\left(
\Psi^{(0)}_{\mathcal{C}}
+
\Psi^{(1)}_{\mathcal{C}}
\right)
=
\begin{pmatrix}
c_{AA} \\
c_{AB} \\
c_{BA} \\
c_{BB}
\end{pmatrix},
\end{equation}
where in the present calculation equal weights $w_{ee} = w_{eh} = w_{he} = w_{hh}$ are adopted as a symmetric benchmark that treats all four quasiparticle sectors on equal footing in the absence of a microscopically derived initial state. To assess the sensitivity of the results to this choice, we verified that the qualitative features of the entropy maps---the position of the crossover in $\Sigma$, the suppression along $p_1=p_2$, and the dependence on $d_1/L$---are preserved when the \textit{ee} sector is weighted twice as heavily as the others, and when a random relative phase is introduced between sectors. The state is normalized as 
\begin{equation}
\Psi_{\text{tot}}
\rightarrow
\frac{\Psi_{\text{tot}}}
{\sqrt{\langle \Psi_{\text{tot}}|\Psi_{\text{tot}}\rangle}}.
\end{equation}

\end{document}

%% file: packages.tex
\usepackage{bm}
\usepackage{graphicx}
\usepackage{units}
\usepackage[english]{babel}
\usepackage{bm}
\usepackage{amsbsy}
\usepackage{amstext}
\usepackage{amssymb}
\usepackage{physics}
\usepackage{units}
\usepackage[utf8]{inputenc}
\DeclareUnicodeCharacter{2060}{}
\usepackage[english]{babel}
\usepackage{wrapfig}
\usepackage{slashed}
\usepackage{bm}
\usepackage{lipsum}

\usepackage{tikz}
\usepackage[compat=1.1.0]{tikz-feynman}
\tikzfeynmanset{warn luatex=false}
\usetikzlibrary{bending}

\usepackage{feynmp-auto}

\usepackage{simpler-wick}
\makeatletter
\def\label#1{\@bsphack
  \begingroup
  \UseHookWithArguments{label}{1}{#1}%
  \protected@write\@auxout{}%
         {\string\newlabel{#1}{{\@currentlabel}{\thepage}%
          {\@currentlabelname}{\@currentHref}{\@kernel@reserved@label@data}}}%
  \endgroup
  \@esphack}
\let\ltx@label\label
\makeatother
\usepackage[
pdfusetitle,
unicode,
colorlinks=true,
linkcolor=blue,
citecolor=blue,
urlcolor=blue
]{hyperref}
\usetikzlibrary{decorations.markings}
\tikzset{W->-/.style={decoration={
			markings,
			mark=at position 0.5*\pgfdecoratedpathlength+2pt with
			{\draw[-latex] (-2pt,0pt) -- (1pt,0pt);}},postaction={decorate}},
	W-<-/.style={decoration={
			markings,
			mark=at position 0.5*\pgfdecoratedpathlength with
			{\draw[latex-] (-2pt,0pt) -- (1pt,0pt);}},postaction={decorate}}
}
\pgfkeys{
	/simplerwick/.cd,
	arrows/.store in=\LstWickArrows,
	arrows/.initial={-,-,-,-,-,-,-,-,-}, 
	arrows={-,-,-,-,-,-,-,-,-},
	positions/.store in=\LstWickPositions,
	positions={+1,+1,+1,+1,+1,+1,+1,+1,+1},
	positions/.initial={+1,+1,+1,+1,+1,+1,+1,+1,+1},
}

\makeatletter
\newcounter{Wick@up}
\newcounter{Wick@down}
\def\swick@end#1#2{
	\swick@setfalse@#1
	\tikzexternaldisable
	\begin{tikzpicture}[remember picture, baseline=(swick-close#1.base)]
		\node[use as bounding box, inner sep=0pt, outer sep=0pt] (swick-close#1) {$\displaystyle #2$};
	\end{tikzpicture}
	\tikz[remember picture, overlay]
	{
		\xdef\myW@style{\empty}
		\foreach \W@X[count=\W@C] in \LstWickArrows
		{\ifnum\W@C=#1
			\xdef\myW@style{\W@X}
			\fi}
		\ifx\myW@style\empty
		\PackageWarning{simpler-wick}{%
			The list arrows has not enough entries!%
		}{}
		\xdef\myW@style{-}
		\fi
		\xdef\myW@pos{-77}
		\foreach \W@X[count=\W@C] in \LstWickPositions
		{\ifnum\W@C=#1
			\xdef\myW@pos{\W@X}
			\fi}
		\ifnum\myW@pos=-77
		\PackageWarning{simpler-wick}{%
			The list positions has not enough entries!%
		}{}
		\xdef\myW@pos{+1}
		\fi
		\ifnum\myW@pos=-1
		\draw[\myW@style] ($(swick-open#1.south) + (0, -3pt)$) 
		-- ($(swick-open#1.base) + (0, -\swick@offset) + \theWick@down*(0, -\swick@sep)$) 
		-- ($(swick-close#1.base) + (0, -\swick@offset) + \theWick@down*(0, -\swick@sep)$) 
		-- ($(swick-close#1.south) + (0, -3pt)$);
		\stepcounter{Wick@down}
		\else
		\stepcounter{Wick@up}
		\draw[\myW@style] ($(swick-open#1.north) + (0, 3pt)$) 
		-- ($(swick-open#1.base) + (0, \swick@offset) + \theWick@up*(0, \swick@sep)$) 
		-- ($(swick-close#1.base) + (0, \swick@offset) + \theWick@up*(0, \swick@sep)$) 
		-- ($(swick-close#1.north) + (0, 3pt)$);
		\fi}
	\tikzexternalenable}
\def\wick@[#1]#2{\setcounter{Wick@up}{0}
	\setcounter{Wick@down}{-1}
	\ifmmode
	\begingroup
	\pgfkeys{
		simplerwick,
		#1}
	\swick@cond@reset
	\swick@count=0
	\def\swick@max{0}
	\def\c{\swick@smart}
	#2
	\dimen0=\swick@sep
	\multiply\dimen0 by \swick@max
	\advance\dimen0 by \swick@offset
	\vbox to \dimen0{}
	\swick@cond@any{
		\PackageWarning{simpler-wick}{%
			I have reached the end of \protect\wick\space with some unclosed
			contractions%
		}{}
	}{}
	\endgroup
	\else
	\PackageWarning{simpler-wick}{%
		\protect\wich\space has been called outside a math environment, this will
		be ignore%
	}
	\fi
}

%% file: authors.tex
\author{Facundo Arreyes}
    \email{facundo.arreyes@uns.edu.ar}
	\affiliation{Departamento de F\'isica, Universidad Nacional del Sur, Av. Alem 1253, B8000, Bah\'ia Blanca, Argentina}
    \affiliation{Instituto de F\'isica del Sur, Conicet, Av. Alem 1253, B8000, Bah\'ia Blanca, Argentina}

\author{Federico Escudero}
\affiliation{IMDEA Nanoscience, Faraday 9, 28049 Madrid, Spain}
    
\author{Arian Gorza}
\affiliation{Departamento de F\'isica, Universidad Nacional del Sur, Av. Alem 1253, B8000, Bah\'ia Blanca, Argentina}
\affiliation{Instituto de F\'isica del Sur, Conicet, Av. Alem 1253, B8000, Bah\'ia Blanca, Argentina}
    
\author{Juan Sebastián Ardenghi}
\affiliation{Departamento de F\'isica, Universidad Nacional del Sur, Av. Alem 1253, B8000, Bah\'ia Blanca, Argentina}
\affiliation{Instituto de F\'isica del Sur, Conicet, Av. Alem 1253, B8000, Bah\'ia Blanca, Argentina}